\documentclass[12pt,a4paper,fleqn]{article}
\usepackage{amsmath,amssymb,amsfonts,amsthm,amsopn,graphicx}
\usepackage{tikz} 
\usetikzlibrary{arrows}
\usepackage{hyperref}
\usepackage{mathrsfs} 

\usepackage[T1,T2A]{fontenc}
\usepackage[utf8]{inputenc}

\allowdisplaybreaks[3]
\newtheorem{theorem}{Theorem}[section]
\newtheorem{definition}[theorem]{Definition}

\newtheorem{proposition}[theorem]{Statement}

\numberwithin{equation}{section} 

\newcommand{\ds}{\displaystyle}
\def\EXP{\textrm{{\large e}}}
\newcommand{\ii}{{\mathtt{i}}}

\newcommand{\bk}{{\textsf{b}}}

\newcommand{\Chi}{\textrm{\Large $\chi$}}

\newcommand{\BH}{\mathsf{H}}

\newcommand{\Vop}{\boldsymbol{V}}

\newcommand{\pol}{\mathsf{P}}

\newcommand{\Cl}{\textrm{\large $\boldsymbol{\gamma}$}}

\newcommand{\hf}{{^1\!\!/\!_2}}

\def\uop{\boldsymbol{u}}
\def\vop{\boldsymbol{v}}

\usepackage[body={15cm,22cm}]{geometry}

\begin{document}

\title{On algebraic structures underlying the rational Kashiwara-Miwa-type models.}
\author{Sergey Sergeev
    \thanks{
   Faculty of Science and Technology, 
   University of Canberra, Bruce ACT 2617, Australia
}
\thanks{ 
Department of Fundamental and Theoretical Physics,
         Research School of Physics and Engineering,
    Australian National University, Canberra, ACT 0200, Australia}
}

\date{}

\maketitle

\begin{abstract}
The rational Kashiwara-Miwa model is an example of an Ising-type integrable model of the statistical physics, related to the six-vertex trigonometric $R$-matrix. Two-spin edge weights of the model are expressed in the terms of $q$-products, its spins are arbitrary integers, and $|q|<1$. We discuss in this paper the algebraic structures underlying the model, in particular  its relation to the $q$-oscillator algebra, to representations of the $q$-oscillator algebra and to the co-product of the $q$-oscillator algebra. 
\end{abstract}

\tableofcontents


\section{Introduction}

There is an integrable model of the lattice statistical physics called the cyclic elliptic Kashiwara-Miwa model \cite{KM}. It belongs to a class of Ising-type models with a two-spin edge interaction and  additional one-spin site weights. Its primary integrability equation is the so-called Onsager's "Star-Triangle relation" \cite{Baxbook}, a specific form of the Yang-Baxter equation for the Ising-type models.

The original Kashiwara-Miwa model is associated with a class of cyclic representations of the elliptic Sklyanin algebra \cite{Skl82}. From the other hand side, there is an object which can be called "the rational Kashiwara-Miwa model", and which is neither elliptic nor cyclic, with $|q|<1$ (this is not elliptic $q$, this is quantum group's $q$, formerly the root of unity for the cyclic representations). This rational model has been derived in \cite{BSKels} as one of various limits of the so-called ``master solution'' of the Star-Triangle relation \cite{Master} (the master solution is also related to the Sklyanin algebra). There is one more limit of the master solution, as well derived in \cite{BSKels} and called "the hyperbolic limit". Its Boltzmann weights are expressed in the terms of Faddeev's quantum dilogarithm \cite{FMod,FK94}, and it also can be seen as a form of the rational Kashiwara-Miwa model.

The algebraic derivation of the Ising-type models, e.g. the chiral Potts model, or the Faddeev-Volkov model, or the master solution, etc., is usually based on a technique of so-called Baxter's vectors\footnote{Baxter used this vectors to derive his $TQ$-equation and to establish a vertex-IRF correspondence \cite{Baxbook,ABF}} \cite{BS90}. The reader can find some modern results in this subject in \cite{descendant,distant}. The present paper has a different motivation. The main motivation here is to make a revision of an algebraic derivation scheme of the rational Kashiwara-Miwa models. By an algebraic scheme we mean an approach based on the Drinfeld-Jimbo quantum group methods, e.g. on the method of Hopf co-products and a representation theory.

A very rough classification scheme for the rational Ising-type models, based on the quantum groups, is the following. It is well known, there are three most useful types of $L$-operators related to the six-vertex $R$-matrix. The first type is the standard $\mathcal{U}_q(\hat{sl}_2)$ $L$-operators, called sometimes the "$L$-operators for the quantum sine-Gordon lattice models" according to the Leningrad group terminology. The corresponding integrable models, including in their Ising-type form the chiral Potts model\footnote{The original Ising model as well as the chiral Potts model are also rational, their algebraic curve parametrisation is the subject of cyclic representations.}, the Faddeev-Volkov model, etc. \cite{BS90,FV95,BMS07,descendant,distant}, are the fundamental examples of the integrable models of statistical physics. The second type of $L$-operators is related to the quantum relativistic Toda chain \cite{Skl85}. They have various quantum mechanical applications.
And the third type of $L$-operators corresponds to the lattice Bose gas \cite{BB92,BS06}. Its algebra of observables is the $q$-oscillator algebra.

This third type $L$-operator and a generalised $q$-oscillator algebra are the main subjects of this paper. The product of $L$-operators defines the co-product of the generalised $q$-oscillator algebra and makes it the Hopf algebra. For majority of representations of the $q$-oscillator algebra there exist their certain form, called the "$\mathbb{V}$-form" in this text, or the  "$\mathbb{V}$-basis". The co-product factorises in such basis, and moreover, the product of $L$-operators is intertwined by a rational Kashiwara-Miwa-type $R$-matrix. The "Star-Triangle" relation can be also described in the terms of co-product. 

We discuss in this paper only three representations of the generalised $q$-oscillator. They are: the Fock space representation providing a novel (but unphysical) solution of the Star-Triangle relation; some reducible generalisation of the standard Fock space representation providing the rational Kashiwara-Miwa model; and one of modular representations -- Faddeev's modular double \cite{FMod,FMod99} -- providing the hyperbolic Kashiwara-Miwa model. As an example, the 3-j symbols are also constructed for the usual Fock space representation. Some other representations are briefly mentioned in Appendix \ref{Add-Section}.

The paper is organised as follows. Firstly, we define the generalised $q$-oscillator algebra, its co-product, and give other required definitions in Section \ref{O-Section}. Then, there is a rather lengthy Section \ref{V-Section} where we define the $\mathbb{V}$-form and consider its properties and application to the co-product in details.
In Section \ref{F-Section} we consider the fundamental Fock space representation, construct its $\mathbb{V}$-form and corresponding solution of the Star-Triangle equation. Then, the reducible $\mathbb{V}$-representation and its irreducible symmetric form providing the rational Kashiwara-Miwa model are considered in Section \ref{F2-Section}. The final section \ref{M-Section} is devoted to the modular representation of the $q$-oscillator algebra related to the Faddeev modular double of Weyl algebra and to the derivation of the hyperbolic Kashiwara-Miwa model.
Some necessary technical details are gathered in 
Appendix Sections \ref{A1-Section},\ref{A2-Section},\ref{Chi-Section}.
Appendix Section \ref{Add-Section} contains a short list of some other representations and 
elements of a "table of multiplication" for them.


\section{Generalised $q$-oscillator and co-product.}\label{O-Section}

Common definitions for the generalised $q$-oscillator algebra and its co-product are given in this section. Also two main useful homomorphisms of the $q$-oscillator algebra to the $q$-Weyl algebra are defined here.

\subsection{Definitions.}

General definitions are collected here.
\begin{definition}
Let the generalised $q$-oscillator algebra is defined by the following relations:
\begin{equation}\label{alg-o}
\mathcal{O}_q[\mathcal{K},\mathcal{K}']\;:\quad \left\{
\begin{array}{l}
\ds 
\mathcal{K} \mathcal{E}^{\pm}\;=\;q^{\pm 1} \mathcal{E}^{\pm} \mathcal{K}\;,\quad
\mathcal{K}' \mathcal{E}^{\pm}\;=\;q^{\pm 1} \mathcal{E}^{\pm} \mathcal{K}'\;,\quad
[\mathcal{K},\mathcal{K}']\;=\;0\;,\\
\\
\ds 
\mathcal{E}^{+}\mathcal{E}^{-}\;=\;1-q^{-1} \mathcal{K}\mathcal{K}'\;,\quad
\mathcal{E}^{-}\mathcal{E}^{+}\;=\;1-q \mathcal{K}\mathcal{K}'\;.
\end{array}\right.
\end{equation}
\end{definition}
One could distinguish the generalised algebra (\ref{alg-o}) and the standard $q$-oscillator algebra  $\mathcal{O}_q\subset \mathcal{O}_q[\mathcal{K},\mathcal{K}']$,
\begin{equation}
\mathcal{O}_q\;:\quad q\mathcal{E}^{+}\mathcal{E}^{-}-q^{-1}\mathcal{E}^{-}\mathcal{E}^{+}\;=\;q-q^{-1}\;.
\end{equation} 
It worth noting,
\begin{equation}
\mathcal{K}\mathcal{K}'\;=\; q\, \biggl( 1 - \mathcal{E}^{+}\mathcal{E}^{-}\biggr)\;\in\;\mathcal{O}_q\;.
\end{equation}

\begin{proposition}
Equations
\begin{equation}\label{coprod}
\begin{array}{ll}
\ds \Delta(\mathcal{E}^-)\;=\;\mathcal{E}^-\otimes\mathcal{E}^- - \mathcal{K}\otimes\mathcal{K}'\;, & 
\ds \Delta(\mathcal{K})\;=\;\mathcal{E}^-\otimes\mathcal{K} + \mathcal{K}\otimes\mathcal{E}^+\;,\\
\\
\ds \Delta(\mathcal{K}')\;=\; \mathcal{K}'\otimes\mathcal{E}^- + \mathcal{E}^+\otimes\mathcal{K}'\;, &
\ds \Delta(\mathcal{E}^+)\;=\; \mathcal{E}^+\otimes\mathcal{E}^+ - \mathcal{ K}'\otimes\mathcal{K}\;
\end{array}
\end{equation}
define a homomorphism $\mathcal{O}_q[\mathcal{K},\mathcal{K}']\;\to\;\mathcal{O}_q[\mathcal{K},\mathcal{K}']\;\otimes\;\mathcal{O}_q[\mathcal{K},\mathcal{K}']$. 
\end{proposition}
\noindent\textbf{Proof} is the direct verification. \hfill $\square$
\\

\noindent
The co-product can be written in the conventional form as the product of $L$-matrices,
\begin{equation}\label{L}
\Delta(L)\;=\;L\;\stackrel{.}{\otimes} \; L\;,\quad 
L\;\stackrel{def}{=}\;\left(\begin{array}{cc} \ds \mathcal{E}^{-} & \ds - \mathcal{K} \\ 
\\
\ds \mathcal{K}' & \ds \mathcal{E}^{+}
\end{array}\right)\;,
\end{equation}
what guarantees the co-associativity.

Let us suppose, there is a representation $\pi_\lambda$ of the algebra $\mathcal{O}_q[\mathcal{K},\mathcal{K}']$, where $\lambda$ is a vector of parameters of the representation. Then, there is a traditional problem to construct the braid group intertwining operator $\hat{S}$,
\begin{equation}\label{RLL}
\pi_{\lambda_1}(L) \, \stackrel{.}{\otimes}\, \pi_{\lambda_2}(L)\;
\check{S}(\lambda_1,\lambda_2)
\;=\;
\check{S}(\lambda_1,\lambda_2)
\;
\pi_{\lambda_2}(L) \,\stackrel{.}{\otimes}\,  \pi_{\lambda_1}(L)\;.
\end{equation}

\subsection{Two homomorphisms}

We will use two special homomorphisms in this paper.

Let $\mathcal{W}_q$ be the universal enveloping of the simple Weyl algebra,
\begin{equation}\label{Weyl}
\mathcal{W}_q\;:\quad \uop\vop\;=\;q\vop\uop\;,\quad q\in\mathbb{C}\;,\quad |q|\;<\;1\;.
\end{equation}
Then the simple homomorphism $\phi: \mathcal{O}_{q}[\mathcal{K},\mathcal{K}']\to \mathcal{W}_q$ is defined by
\begin{equation}\label{Hom1}
\phi(\mathcal{K}) = \omega\, \uop,\;\; 
\phi(\mathcal{K}') = \omega^{-1}\uop,\;\;
\phi(\mathcal{E}^{+}) = f(\uop) \, \vop,\;\; 
\phi(\mathcal{E}^{-}) = \vop^{-1} g(\uop),
\end{equation}
where
\begin{equation}\label{fg}
f(\uop)g(\uop)\;=\;1-q^{-1}\uop^2\;,\quad \omega\;\in\;\mathbb{C}\;.
\end{equation}
However, there is another remarkable homomorphism $\pi_{\omega,\mu}: \mathcal{O}_{q}[\mathcal{K},\mathcal{K}']\to \mathcal{W}_q$,
\begin{equation}\label{Hom2}
\left\{
\begin{array}{l}
\ds 
\omega^{-1}\pi_{\omega,\mu}(\mathcal{K})\;=\;
\omega\, \pi_{\omega,\mu}(\mathcal{K}')\;=\;(\boldsymbol{v} - \boldsymbol{v}^{-1})\,(\boldsymbol{u}-\boldsymbol{u}^{-1})^{-1}\;,\\
\\
\ds 
\pi_{\omega,\mu}(\mathcal{E}^{+})\;=\;\mu^{-1}\; (\boldsymbol{v}^{-1}\boldsymbol{u}-\boldsymbol{v}\boldsymbol{u}^{-1})\,(\boldsymbol{u}-\boldsymbol{u}^{-1})^{-1}\;,\\
\\
\ds
\pi_{\omega,\mu}(\mathcal{E}^{-})\;=\;\mu\; (\boldsymbol{v}\boldsymbol{u}-\boldsymbol{v}^{-1}\boldsymbol{u}^{-1})\,
(\boldsymbol{u}-\boldsymbol{u}^{-1})^{-1}\;.
\end{array}\right.
\end{equation}
Here $\omega$ and $\mu$ are the "parameters of a representation" (parameter $\mu$ for the homomorphism (\ref{Hom1}) is hidden inside $f(u)$ and $g(u)$).
It will be convenient to use one more form of (\ref{Hom2}), namely $\pi'_{\omega,\mu}$ defined as
\begin{equation}
\pi'_{\omega,\mu}(\star)\;=\; (\uop-\uop^{})^{-1} \, \pi_{\omega,\mu}^{}(\star)\, (\uop-\uop^{-1})\;.
\end{equation}
Explicitly, $\pi'_{\omega,\mu}$ is given by
\begin{equation}\label{Hom2-2}
\left\{
\begin{array}{l}
\ds 
\omega^{-1}\pi_{\omega,\mu}'(\mathcal{K})\;=\;
\omega\, \pi_{\omega,\mu}(\mathcal{K}')\;=\;(\boldsymbol{u}-\boldsymbol{u}^{-1})^{-1}\,(\boldsymbol{v} - \boldsymbol{v}^{-1})\;,\\
\\
\ds 
\pi_{\omega,\mu}'(\mathcal{E}^{+})\;=\;\frac{q}{\mu}\;
(\boldsymbol{u}-\boldsymbol{u}^{-1})^{-1}\, (\boldsymbol{u}\boldsymbol{v}^{-1}-
\boldsymbol{u}^{-1}\boldsymbol{v})\;,\\
\\
\ds
\pi_{\omega,\mu}'(\mathcal{E}^{-})\;=\;\frac{\mu}{q}\;
(\boldsymbol{u}-\boldsymbol{u}^{-1})^{-1}\, (\boldsymbol{u}\boldsymbol{v}-
\boldsymbol{u}^{-1}\boldsymbol{v}^{-1})
\;.
\end{array}\right.
\end{equation}

\subsection{Relation to the six-vertex $R$-matrix.}

Before we proceed with the generalised $q$-oscillator, it worth to clarify how matrix $L$, eq. (\ref{L}), is related to the six-vertex $R$-matrix.

Let us write the six-vertex $R$-matrix as follows:
\begin{equation}
R(\lambda,\mu)\;=\;
\left(\begin{array}{cccc}
[q\mu/\lambda] & 0 & 0 & 0 \\
0 & \ds [\mu/\lambda] &  [q] & 0 \\
0 & \ds [q] & [\mu/\lambda] & 0 \\
0 & 0 & 0 & [q\mu/\lambda]
\end{array}
\right)\;.
\end{equation}
If 
\begin{equation}\label{Lop}
L(\lambda) \;=\; \left(
\begin{array}{cc}
\ds \mathcal{E}^{-} & \ds -\mathcal{K}/\lambda \\ \ds \lambda\mathcal{K}' & \mathcal{E}^{+}
\end{array}\right)\;=\;
\left(\begin{array}{cc} 1 & 0 \\ 0 & \lambda\end{array}\right)\; L \; 
\left(\begin{array}{cc} 1 & 0 \\ 0 & \lambda\end{array}\right)^{-1}
\;,
\end{equation}
then there is the usual $RLL$ relation:
\begin{equation}
\sum_{j_1,j_2}\; R(\lambda,\mu)_{i_1,i_2}^{\;\; j_1,j_2} \, L(\lambda)_{j_1}^{\;\; k_1}\; L(\mu)_{j_2}^{\;\; k_2}\;=\;
\sum_{j_1,j_2}\; L(\mu)_{i_2}^{\;\; j_2}\; L(\lambda)_{i_1}^{\;\; j_1}\; R(\lambda,\mu)_{j_1,j_2}^{\;\; k_1,k_2}\;.
\end{equation}
Due to the symmetry of the six-vertex $R$-matrix, the same $RLL$ relation is valid for 
$L(\lambda)\sigma_x$, $\sigma_x L(\lambda)$ and $\sigma_xL(\lambda)\sigma_x$, where $\sigma_x$ is the first Pauli matrix. The matrix (\ref{Lop}) is useless in the theory of the integrable systems since it depends on the spectral parameter $\lambda$ in the trivial way. However, $L$-operator $L(\lambda)\sigma_x$ is useful, it defines the integrable lattice Bose-gas..

\section{Co-product and its $\mathbb{V}$-form.}\label{V-Section}

For any representation of the algebra $\mathcal{O}_q[\mathcal{K},\mathcal{K}']$ with invertible $\mathcal{K}$ and $\mathcal{K}'$ there exists an universal basis, which we will call "the $\mathbb{V}$-form of the representation'', in which the Klebsh-Gordan coefficients become trivial.

\subsection{The origin of the $\mathbb{V}$-form.}

According to its definition, the algebra $\mathcal{O}_q[\mathcal{K},\mathcal{K'}]$ with invertible $\mathcal{K}$ and $\mathcal{K}'$ has a center, $\ds \omega^2=\frac{\mathcal{K}}{\mathcal{K}'}$. For the irreducible representations $\omega\in\mathbb{C}$. Then for a representation with fixed $\omega$ let
\begin{equation}
\omega^{-1}\,\mathcal{K}\;=\;\omega\;\mathcal{K}'\;=\;\mathcal{K}_0
\end{equation}
Here we do not use a symbol of representation like $\pi_{\omega,\mu}(\mathcal{K})$ since a representation is far from a definition.

Next, we consider the product of two representations, such that the first one has the parameter $\omega_1$, and the second one has the parameter $\omega_2$. Then, we can combine the co-product relations and rewrite the $\omega$-condition for the co-product as the null space of the following operator,
\begin{equation}\label{HH}
\omega^{-1}\, \Delta(\mathcal{K}) - \omega\, \Delta(\mathcal{K}') 
\;\equiv\;\mathcal{K}_0\otimes\mathcal{K}_0 \, \biggl( \mathcal{H}(\frac{\omega}{\omega_2}) \otimes 1 \,-\, 1\otimes \mathcal{H}(\frac{\omega_1}{\omega})\biggr)\;,
\end{equation}
where
\begin{equation}\label{H}
\mathcal{H}(\mu)\;\stackrel{def}{=}\;\mathcal{K}_0^{-1}\, \biggl(\mu^{-1}\mathcal{E}^{-} - \mu\mathcal{E}^{+}\biggr)\;.
\end{equation}
This observation gives rise of the 
\begin{definition}\label{VDEF}
$\mathbb{V}[\mu]$-form of any representation, or its $\mathbb{V}[\mu]$-basis, is the system of the states $|\omega;\mu,\xi\rangle$ such that firstly
\begin{equation}
\biggl(\omega^{-1}\mathcal{K}-\omega\mathcal{K}'\biggr)\; |\omega;\mu,\xi\rangle \;=\;0\;,
\end{equation}
secondly
\begin{equation}\label{Vdef}
\mathcal{H}(\mu) \; |\omega;\mu,\xi\rangle \;=\; |\omega;\mu,\xi\rangle \; (\xi+\xi^{-1})\;,
\end{equation}
and thirdly, $\xi\in\mathbb{S}$ for some set $\mathbb{S}$ such that the system of the states
\begin{equation}\label{V}
\mathbb{V}\;=\;\textrm{Span}\biggl\{|\omega,\mu,\xi\rangle,\;\; \xi\in\mathbb{S}\biggr\}
\end{equation}
is complete in the space of the initial representation.
\end{definition}

\subsection{General properties of $\mathbb{V}$-basis.}

\begin{proposition}
Let the states $|\omega;\mu,\xi\rangle$ satisfy relation (\ref{Vdef}). Then they obey the additional set of relations,
\begin{equation}\label{Br}
\left\{
\begin{array}{l}
\biggl(\mu^{-1}\mathcal{E}^{-} - \mathcal{K}_0 \xi \biggr) |\mu,\xi\rangle \;=\;
\biggl(\mu \mathcal{E}^{+} + \mathcal{K}_0 \xi^{-1} \biggr) |\mu,\xi\rangle \;=\; |\mu,q^{-1}\xi\rangle \, A(\xi)\;,\\
\\
\ds \biggl(\mu^{-1}\mathcal{E}^{-} - \mathcal{K}_0 \xi^{-1}\biggr) |\mu,\xi\rangle \;=\;
\biggl(\mu \mathcal{E}^{+} + \mathcal{K}_0 \xi \biggr) |\mu,\xi\rangle \;=\; |\mu,q\xi\rangle \, A(q\xi)^{-1}\;,
\end{array}\right.
\end{equation}
where we omit the argument $\omega$ in the state notations for the shortness, and the single undefined function $A(\xi)$ is a subject of a normalisation of $|\omega;\mu,\xi\rangle$.
\end{proposition}
\noindent\textbf{Proof:} The left equalities in the equations (\ref{Br}) are simple consequences of (\ref{Vdef}). To prove the right equalities, we define 
\begin{equation}
\mathcal{G}^{-}(\mu,\xi )\;=\; \mu^{-1} \, \mathcal{E}^{-} - \mathcal{K}_0 \, \xi\;,\quad
\mathcal{G}^{+}(\mu,\xi) \;=\; \mu \, \mathcal{E}^{+} + \mathcal{K}_0 \, \xi\;.
\end{equation}
These operators have the straightforward properties,
\begin{equation}
\begin{array}{l}
\ds \mathcal{G}^{+}(\mu,q^{-1}\xi) \; \mathcal{G}^{-}(\mu,\xi) \;=\; 
1+q^{-1} \mathcal{K}_0\xi\, \biggl( \mathcal{H}(\mu) - \xi - \xi^{-1}\biggr)\;,\\
\\
\ds \mathcal{G}^{-}(\mu,q\xi) \; \mathcal{G}^{+}(\mu,\xi) \;=\; 
1 + q \mathcal{K}_0\xi \,
\biggr( \mathcal{H}(\mu) - \xi - \xi^{-1} \biggl)\;.
\end{array}
\end{equation}
It follows,
\begin{equation}
\mathcal{G}^{+}(\mu,q^{-1}\xi) \, \mathcal{G}^{-}(\mu,\xi) \, |\mu,\xi\rangle \;=\; |\mu,\xi\rangle\;,\quad
\mathcal{G}^{-}(\mu,q\xi) \, \mathcal{G}^{+}(\mu,\xi) \, |\mu,\xi\rangle \;=\; |\mu,\xi\rangle\;,
\end{equation}
what provides the right equalities in the formulas (\ref{Br}). \hfill $\square$

The relations (\ref{Br}) have the elementary consequences:
\begin{equation}\label{X1}
\left\{
\begin{array}{l}
\ds \mathcal{K}_0 \, |\mu,\xi\rangle \;=\; \biggl( |\mu,q\xi\rangle A(q\xi)^{-1} - |\mu,q^{-1}\xi\rangle A(\xi)\biggr) / (\xi-\xi^{-1})\;,\\
\\
\ds \mathcal{E}^{+} |\mu, \xi\rangle \;=\; \mu^{-1}\biggl( |\mu,q^{-1}\xi\rangle \xi A(\xi) - |\mu,q\xi\rangle \xi^{-1} A(q\xi)^{-1}\biggr) / (\xi-\xi^{-1})\;,\\
\\
\ds \mathcal{E}^{-} |\mu,\xi\rangle \;=\; \mu\, \biggl( |\mu,q\xi\rangle \xi A(q\xi)^{-1} - |\mu, q^{-1}\xi\rangle \xi^{-1} A(\xi)\biggr) / (\xi-\xi^{-1})\;.
\end{array}\right.
\end{equation}
These correspond to the representation of the $q$-oscillator (\ref{alg-o}) via the homomorphism (\ref{Hom2}) with 
\begin{equation}\label{W2rep}
\uop |\mu,\xi\rangle \;=\; |\mu,\xi\rangle\, \xi\;,\quad 
\vop |\mu,\xi\rangle \;=\; |\mu,q\xi\rangle \, A(q\xi)^{-1}\;.
\end{equation}
Let us note, one has straightforwardly in the terms of the homomorphism (\ref{Hom2}) 
\begin{equation}
\mathcal{H}(\mu)=\uop+\uop^{-1},\;\;
\mu^{-1}\mathcal{E}^{-}-\mathcal{K}_0\uop=\vop^{-1},\;\;
\mu^{-1}\mathcal{E}^{-}-\mathcal{K}_0\uop^{-1}=\vop\;.
\end{equation}

\subsection{Co-product and $\mathbb{V}$-form.}

The $\mathbb{V}$-basis have the following advantage when one considers the co–product. Let the double-state $|\!|\xi\rangle$ is defined by
\begin{equation}\label{X00}
|\!|\xi\rangle \;=\; |\omega_1;\frac{\omega}{\omega_2},\xi\rangle \otimes |\omega_1;\frac{\omega_1}{\omega},\xi\rangle\;.
\end{equation}
Then, according to (\ref{X1}) and by means of elementary algebraic manipulations one obtains
\begin{equation}\label{X3}
\left\{
\begin{array}{l}
\ds \Delta(\mathcal{K}) \, |\!|\xi\rangle \;=\; \omega \biggl( |\!|q\xi\rangle A(q\xi)^{-2} - 
|\!| q^{-1}\xi\rangle A(\xi)^2\biggr) / (\xi-\xi^{-1})\;,\\
\\
\ds \Delta(\mathcal{K}') \, |\!|\xi\rangle \;=\; \omega^{-1} \biggl( |\!|q\xi\rangle A(q\xi)^{-2} - 
|\!| q^{-1}\xi\rangle A(\xi)^2\biggr) / (\xi-\xi^{-1})\;,\\
\\
\ds \Delta(\mathcal{E}^{+})\, |\!|\xi\rangle \;=\; \frac{\omega_2}{\omega_1}\,
\biggl( |\!| q^{-1}\xi \rangle \xi A(\xi)^2 - |\!| q\xi \rangle \xi^{-1} A(q\xi)^{-2}\biggr) / (\xi-\xi^{-1})\;,\\
\\
\ds \Delta(\mathcal{E}^{-})\,  |\!| \xi \rangle \;=\; \frac{\omega_1}{\omega_2}\,
\biggl( |\!| q\xi \rangle \xi A(q\xi)^{-2} - |\!| q^{-1}\xi \rangle \xi^{-1} A(\xi)^2\biggr) / (\xi-\xi^{-1})\;,
\end{array}\right.
\end{equation}
In a plain language it means, all unwanted terms identically cancel from (\ref{X3}).
Evidently, this is again the $\mathbb{V}$-form similar to (\ref{X1}) up to the change $A(\xi)\to A(\xi)^2$. In particular, if $A(\xi)=1$, then one can make the evident conclusion:
\begin{equation}\label{X01}
|\!|\omega;\frac{\omega_1}{\omega_2},\xi\rangle\;=\;
|\omega_1;\frac{\omega}{\omega_2},\xi\rangle \otimes |\omega_2;\frac{\omega_1}{\omega},\xi\rangle \;,
\end{equation}
where all notations are written in accordance to the definition \ref{VDEF}. 
Also, one can define a "Klebsh-Gordan projection symbol" $\Cl$ in order to make the relation (\ref{X01}) more formal:
\begin{equation}\label{X01-2}
\Cl\; 
|\omega;\frac{\omega_1}{\omega_2},\xi\rangle\;=\;
|\omega_1;\frac{\omega}{\omega_2},\xi\rangle \otimes |\omega_2;\frac{\omega_1}{\omega},\xi\rangle \;.
\end{equation}
This symbol allows one to rewrite (\ref{X01}) in the form\footnote{On the language of the homomorphism (\ref{Hom2}), this is 
$$
\vop\otimes \vop \; \Cl\;=\;\Cl\; \vop\;,\quad
\uop\otimes 1\; \Cl\;=\; 1\otimes\uop\; \Cl\;=\; \Cl\; \uop\;.
$$
}
\begin{equation}
\pi_{\omega_1,\omega/\omega_2}(L) 
\,\stackrel{.}{\otimes}\,
\pi_{\omega_2,\omega_1/\omega} (L)\;\Cl
\;=\;
\Cl\; \pi_{\omega,\omega_1/\omega_2}(L)\;.
\end{equation}
One could note that $\pi_{\omega,\mu}$ so far is just an idea of a representation since we did not specify the nature of the set $\mathbb{S}$ in the definition (\ref{V}), and we did not specify symmetry properties, asymptotic behaviour etc. -- we did not specify everything defining an irreducible representation.

\subsection{Dual basis for $\mathbb{V}$-form.}

The set $|\mu,\xi\rangle$ has been arisen and the set of the ``right eigenvectors'' of the operator 
$\mathcal{H}(\mu)$, eq. (\ref{Vdef}). In the same time one can define the set of its 
``left eigenvectors'':
\begin{equation}\label{X4}
\langle \mu,\xi |  \; \mathcal{H}(\mu)\;=\;(\xi+\xi^{-1}) \, \langle \mu,\xi|\;.
\end{equation}
A normalisation must be introduced carefully. It is convenient to fix the normalisation of the left and of the right eigenvectors as follows:
\begin{equation}\label{X5}
\langle \mu,\xi| \mu,\xi'\rangle \;=\;\frac{\delta(\xi,\xi')}{S(\xi)}\;,
\end{equation}
where $\delta(\xi,\xi')$ is a properly defined delta-function on the set $\mathbb{S}$, and the function $S(xi)$ could satisfy 
\begin{equation}\label{S1}
S(q^{-1}\xi)\;=\;\frac{[q^{-1}\xi]}{[\xi]}\; \frac{A(\xi)}{B(\xi)} \; S(\xi)\;.
\end{equation}
According to the normalisation (\ref{X5},\ref{S1}), the algebra acts on the co-states as follows\footnote{The analogue of (\ref{Br}) is
\begin{equation}
\langle \mu, \xi | (\frac{q}{\mu} \mathcal{E}^{-}-\xi^{-1}\mathcal{K}_0) = B(\xi)^{-1} \langle \mu, q^{-1}\xi |,\quad
\langle \mu, \xi | (\frac{q}{\mu} \mathcal{E}^{-}-\xi \mathcal{K}_0) = B(q\xi) \langle \mu, q \xi |.
\end{equation}
}:
\begin{equation}\label{X1-2}
\left\{
\begin{array}{l}
\ds 
\langle \mu,\xi | \, \mathcal{K}_0 \, \;=\; \frac{1}{\xi-\xi^{-1}}\,\biggl( B(\xi)^{-1} \langle \mu,q^{-1}\xi |
 - B(q\xi)\, \langle \mu, q\xi |\biggr)\;,\\
\\
\ds 
\langle \mu,\xi | \, \mathcal{E}^{+} \;=\; 
\frac{q/\mu}{\xi-\xi^{-1}}\, \biggl( B(q\xi) \xi \langle \mu,q\xi| 
 - B(\xi)^{-1}\xi^{-1}\langle \mu,q^{-1} \xi | \biggr) \;,\\
\\
\ds 
\langle \mu, \xi | \, \mathcal{E}^{-} \;=\; 
\frac{\mu/q}{\xi-\xi^{-1}}\, 
\biggl( B(\xi)^{-1} \xi \langle \mu,q^{-1}\xi | 
 - B(q\xi) \xi^{-1} \langle \mu, q\xi | \biggr) \;.
\end{array}\right.
\end{equation}
This corresponds to the form (\ref{Hom2-2}) of the homomorphism (\ref{Hom2}) with  
\begin{equation}
\langle \mu,\xi | \,\uop\;=\; \xi\, \langle \mu,\xi|\;,\quad 
\langle \mu,\xi | \, \vop\;=\; B(\xi)^{-1}\, \langle \mu,q^{-1}\xi|\;.
\end{equation} 
What is the actual meaning of all these formulas. One solves the ``eigenvector'' equations (\ref{Vdef}) and (\ref{X4}) and constructs the ``eigenvectors'' in a most convenient way. Then one looks at the relations (\ref{Br}) and their left analogue and fixes the values of $A(\xi)$ and $B(\xi)$. And only then one computes the normalisation factor (\ref{X5}) of the states obtained, and this normalisation factor must obey (\ref{S1}) identically.

On this stage, before going into details of particular representations, we fix the co-module $\mathbb{V}$ by means of (\ref{X5},\ref{S1}) and postulate the completeness,
\begin{equation}
1\;=\;\sum_{\xi\in\mathbb{S}} |\mu,\xi\rangle \, S(\xi)\, \langle \mu,\xi|\;.
\end{equation}

However, a construction of the co-module for the co-product (\ref{X00}) is not as straightforward as before. In order to repeat the steps (\ref{HH}), (\ref{H}), (\ref{Br}) and so on for the co-module, one could note that the equation (\ref{HH}) can be re-written identically as
\begin{equation}
\omega^{-1}\, \Delta(\mathcal{K}) - \omega\, \Delta(\mathcal{K}') 
\;\equiv\; \biggl( \mathcal{H}'(\frac{\omega}{\omega_2}) \otimes 1 \,-\, 1\otimes \mathcal{H}'(\frac{\omega_1}{\omega})\biggr)\,\mathcal{K}_0\otimes\mathcal{K}_0 \;,
\end{equation}
where
\begin{equation}
\mathcal{H}'(\mu) \;=\; \biggl( \mu^{-1}\mathcal{E}^{-} - \mu \mathcal{E}^{+}\biggr)\; \mathcal{K}_0^{-1}\;=\;\mathcal{K}_0^{}\, \mathcal{H}(\mu) \, \mathcal{K}_0^{-1}
\;=\;\mathcal{H}(q\mu)\;.
\end{equation}
It follows, the left eigenvectors of the operator $\mathcal{H}'(\mu)$ are required for the construction of the co-module. They are in our notations
\begin{equation}\label{qmu}
\langle q\mu,\xi| \;\sim\; \langle \mu,\xi|\, \mathcal{K}_0^{-1}\;:\quad
\langle q\mu,\xi| \, \mathcal{H}'(\mu) \;=\; (\xi+\xi^{-1})\, \langle q\mu,\xi|\;.
\end{equation}
The representation of the algebra on the states $\langle q\mu,\xi |$ is given by the formula (\ref{X1-2}) with the change $\mu\to q\mu$. 
Repeating now the reasoning giving the relation between (\ref{X1}) and (\ref{X3}), one comes to the following form of the co-module of the co-product:
\begin{equation}\label{X02}
\langle \omega_1; q\frac{\omega}{\omega_2},\xi | \otimes \langle \omega_2; q\frac{\omega_1}{\omega},\xi | \;=\; \langle \omega; q \frac{\omega_1}{\omega_2},\xi |\!|\;,
\end{equation}
with the same comments as to the formula (\ref{X01}), i.e. with the change $B(\xi)\to B(\xi)^2$. 
Following ideas of (\ref{X01-2}), it is convenient to introduce another projection symbol $\Cl^{*}$,
\begin{equation}\label{X02-2}
\langle \omega; q \frac{\omega_1}{\omega_2},\xi | \, \Cl^*\;=\;
\langle \omega_1; q\frac{\omega}{\omega_2},\xi | \otimes \langle \omega_2; q\frac{\omega_1}{\omega},\xi |\;,
\end{equation}
so that 
\begin{equation}
\Cl^{*}\; \pi_{\omega_1,q\frac{\omega}{\omega_2}}(L)\,
\stackrel{.}{\otimes}\,
\pi_{\omega_2,q\frac{\omega_1}{\omega}}(L)\;=\;
\pi_{\omega,q\frac{\omega_1}{\omega_2}}(L)\;
\Cl^*\;.
\end{equation}

\subsection{Co-product of equivalent representations}

Now one can turn to the question of the normalisation of the states $|\!|\xi\rangle$ and $\langle \xi |\!|$ in the co-product. According to (\ref{qmu}), one can define 
\begin{equation}
\langle \omega; \frac{\omega_1}{\omega_2},\xi |\!|\;\sim\;
\langle \omega; q\frac{\omega_1}{\omega_2},\xi | \, \Delta(\mathcal{K})\;,
\end{equation}
so that one expects 
\begin{equation}
\langle \omega;\frac{\omega_1}{\omega_2},\xi |\!| \omega';\frac{\omega_1}{\omega_2},\xi'\rangle \;=\;
\frac{\delta(\omega,\omega') \delta(\xi,\xi')}{S_2(x)}\;,
\end{equation}
where $S_2$ corresponds to (\ref{S1}) up to the change $A\to A^2$ and $B\to B^2$. 
But, a condition for the emerging of delta-function for $\omega$ and $\omega'$ is the condition defining a set $\Omega$, $\omega,\omega'\in\Omega$. After this, we can write the ``table of multiplication'' in the form 
\begin{equation}\label{VOV}
\mathbb{V}\otimes\mathbb{V}\;=\;\Omega\otimes\mathbb{V}\;.
\end{equation}
In what follows, giving a ``table of multiplication'' for particular representations, we will imply this order in notations: a set $\Omega$ followed by a representation $\mathbb{V}$ in the right hand side of a ``multiplication formula''.

\subsection{Intertwining operator $\hat{V}_{\mu,\mu'}$}

We noted above, the $\mathbb{V}$-form can be seen as an intermediate step of a definition of a representation, related to the homomorphism (\ref{Hom2}),
\begin{equation}
\mathcal{O}_q[\mathcal{K},\mathcal{K}']\;\;
\stackrel{\pi_{\omega,\mu}}{\to}\;\; \mathcal{W}_q\;\;
\to \;\; \biggl\{|\xi\rangle\;:\;\; \uop\, |\xi\rangle = |\xi\rangle \, \xi\biggr\}\;\;
\stackrel{\mathbb{S}}{\to}\;\;\cdots
\end{equation}
where the dots stand for a specification of $\xi\in\mathbb{S}$, additional conditions, etc.

The representations with different values of the parameter $\mu$ are equivalent. The similarity operator
\begin{equation}\label{Vpi}
\hat{V}_{\mu,\mu'} \pi_{\omega,\mu'}(\star)\;=\;\pi_{\omega,\mu}(\star) 
\hat{V}_{\mu,\mu'}\;,
\end{equation}
can be presented as
\begin{equation}\label{V-1}
\hat{V}_{\mu,\mu'}\;=\;\left(\frac{\mu}{\mu'}\right)^{-\mathcal{N}}\;,\quad \textrm{where}\quad \mathcal{K}\mathcal{K}'\;=\;q^{2\mathcal{N}}\;.
\end{equation}
It worth noting that this operator in general is related to the standard $q$-oscillator algebra $\mathcal{O}_q$ rather than to the generalised algebra $\mathcal{O}_q[\mathcal{K},\mathcal{K}']$.
The form of the matrix elements of the operator $\hat{V}_{\mu,\mu'}$ follows from (\ref{Vpi}), 
\begin{equation}\label{Vpi2}
\langle \xi | \; \hat{V}_{\mu,\mu'} \; \pi_{\mu'}(\star) \; | \xi'\rangle \;=\;
\langle \xi | \; \pi_{\mu}(\star) \; \hat{V}_{\mu,\mu'} \; | \xi' \rangle\;,
\end{equation}
and in the terms of the $\mathbb{V}$-form they can be written as
\begin{equation}\label{V-2}
\langle \xi | \hat{V}_{\mu,\mu'} | \xi'\rangle \;\stackrel{def}{=}\;
V_{\mu/\mu'}(\xi,\xi')\;=\; 
\langle \mu,\xi | \mu',\xi'\rangle\;.
\end{equation}
The consequence of (\ref{Vpi2},\ref{X1},\ref{X5}) are the recursion relations
\begin{equation}\label{req}
\left\{
\begin{array}{l}
\ds 
\frac{V_{\mu/\mu'}(q^{-1}\xi,\xi')}{V_{\mu/\mu'}(\xi,q^{-1}\xi')} \;=\; 
\biggl( -B(\xi)A(\xi') \biggr) \,\cdot\,\frac{\xi'-\xi q\mu'/\mu}
{\xi-\xi' q\mu'/\mu}\;,\\
\\
\ds 
\frac{V_{\mu/\mu'}(\xi,q\xi')}{V_{\mu/\mu'}(q^{-1}\xi,\xi')} =
\left(-\frac{A(q\xi')}{B(\xi)}\right) \,\cdot\, \frac{\xi\xi'-q\mu'/\mu}{1- \xi\xi' q\mu'/\mu}.
\end{array}\right.
\end{equation}
These equations could be solved only when we will fix a representation, fix $\mathbb{S},A,B$ and fix a class of functions. So far the relations (\ref{req}) are just a guideline.
Nevertheless, \`a priori one expects the transitivity, 
\begin{equation}
\hat{V}_{\mu,\mu'} \; \hat{V}_{\mu',\mu''}\;=\;\hat{V}_{\mu,\mu''}\;,\quad 
\hat{V}_{\mu,\mu}\;=\;1\;,
\end{equation}
what in matrix elements becomes 
\begin{equation}
\sum_{\xi'\in\mathbb{S}} V_{\mu/\mu'}(\xi,\xi') \, S(\xi')\, V_{\mu'/\mu''}(\xi',\xi'')\;=\;
V_{\mu/\mu''}(\xi,\xi'')\;,\quad V_1(\xi,\xi')\;=\;\frac{\delta(\xi,\xi')}{S(\xi)}\;.
\end{equation}
One could mention in advance, this two-point function, $V_{\mu/\mu'}(\xi,\xi')$, is the Boltzmann weight for a rational or for a hyperbolic Kashiwara-Miwa model.

\subsection{Intertwining operator $\hat{\overline{V}}_z$}

There is one more intertwining operator with a completely different meaning. Namely, 
relation
\begin{equation}\label{oV1}
\pi_{\omega_1,\mu_1}\otimes \pi_{\omega_2,\mu_2} \biggl(\Delta(\star)\biggr)\;
\hat{\overline{V}}_z\;=\;
\hat{\overline{V}}_z\; 
\pi_{\omega_1',\mu_1'}\otimes \pi_{\omega_2',\mu_2'} \biggl(\Delta(\star)\biggr)
\end{equation}
where $\hat{\overline{V}}_z$ is a diagonal operator,
\begin{equation}\label{oV2}
\hat{\overline{V}}_x\; |\xi_1\rangle \otimes |\xi_2\rangle \;=\;
|\xi_1\rangle \otimes |\xi_2\rangle \; \overline{V}_z(\xi_1,\xi_2)\;,
\end{equation}
is satisfied if there are the following relations between the parameters,
\begin{equation}\label{oV3}
\omega_1'\;=\;\omega_1\frac{\mu_1}{\mu_1'}\;,\quad
\omega_2'\;=\;\omega_2\frac{\mu_2'}{\mu_2}\;,\quad
\mu_1'\mu_2'\;=\;\varepsilon q \frac{\omega_1}{\omega_2}\;,\quad
\varepsilon^2\;=\;1\;,
\end{equation}
and the function $\overline{V}_z(\xi_1,\xi_2) $ satisfies the difference relations\footnote{Additional conditions could be 
\begin{equation}\label{oVs}
\overline{V}_z(\xi_1,\xi_2)\;=\;\overline{V}_z(\xi_2,\xi_1)\quad \textrm{и}\quad
\overline{V}_z(\xi_1,\xi_2)\;\overline{V}_{1/z}(\xi_1,\xi_2)\;=\;1\;.
\end{equation}
}
\begin{equation}\label{oV4}
\frac{\overline{V}_z(\xi_1,q\xi_2)}{\overline{V}_z(q^{-1}\xi_1,\xi_2)}\;=\;
-\varepsilon \; \frac{\xi_1\xi_2-\varepsilon z}{1-\xi_1\xi_2\varepsilon z}\;,\quad
\frac{\overline{V}_z(q^{-1}\xi_1,\xi_2)}{\overline{V}_z(\xi_1,q^{-1}\xi_2)}\;=\;
-\varepsilon \; \frac{\xi_2-\xi_1\varepsilon z}{\xi_1-\xi_2\varepsilon z}\;,
\end{equation}
and parameter $z$ is defined as
\begin{equation}\label{oV5}
z\;=\;\frac{\mu_1'\mu_2'}{\mu_1\mu_2}\;=\;\varepsilon \frac{q}{\mu_1\mu_2} \frac{\omega_1}{\omega_2}\;.
\end{equation}
There are only three relations in (\ref{oV3}). This is because there are only three independent parameters in the co-product of $\mathbb{V}$-forms.

The relations (\ref{oV3}) look clumsy. To improve them, it is convenient to re-define the parameters of $\mathbb{V}$-form. Namely, let 
\begin{equation}\label{oV6}
\pi_{\omega,\mu}\;\equiv\; \pi_{(x,x')}\;,\quad x\;=\;\lambda\omega\;,\;\; x'\;=\;
\omega/\lambda\;.
\end{equation}
Then the relation (\ref{oV1}) with the extra requirements (\ref{oV3}) is simplified,
\begin{equation}\label{oV8}
\pi_{(x,x')}\otimes \pi_{(y,y')} \biggl(\Delta(\star)\biggr)\;
\hat{\overline{V}}_{z}\;=\;
\hat{\overline{V}}_{z}\; 
\pi_{(x,x'/z)}\otimes \pi_{(zy,y')} \biggl(\Delta(\star)\biggr)\;,\quad 
z\;=\;\varepsilon q \frac{x'}{y}\;.
\end{equation}
In the terms of the parametrisation (\ref{oV6}) the relation (\ref{Vpi}) can be rewritten as
\begin{equation}\label{oV9}
\pi_{(x,x')}(\star)\; \hat{V}_{\lambda}\;=\;
\hat{V}_{\lambda}\;\pi_{(x/\lambda,\lambda x')}(\star)\;.
\end{equation}
One could note the evident correspondence between the functions $V_z(\xi,\xi')$ (\ref{req}) and  $\overline{V}_z(\xi,\xi')$ (\ref{oV4}). In what follows, we will see two particular cases,
\begin{equation}
\begin{array}{l}
\ds A=1\;,\;\; B=-1\;,\;\; \varepsilon=-1\;\;\Rightarrow\;\; \overline{V}_z(\xi,\xi')\;\sim\;V_{-q/z}(\xi,\xi')\;,\\
\\
\ds A=1\;,\;\;B=1\;,\;\; \varepsilon=1\;,\;\;\Rightarrow\;\; \overline{V}_z(\xi,\xi')\;\sim\; V_{q/z}(\xi,\xi')\;.
\end{array}
\end{equation}

\subsection{Intertwining operator for the co-product in the terms of 3-j symbols.}

 Now we can come back to the initial problem -- construction of the intertwining operator $\check{S}$, eq. (\ref{RLL}),
\begin{equation}\label{SLL}
\pi_{\omega_1,\mu_1}(L) \, \stackrel{.}{\otimes}\, \pi_{\omega_2,\mu_2}(L)
\;\check{S}
\;=\;
\check{S}\; 
\pi_{\omega_2,\mu_2}(L) \, \stackrel{.}{\otimes}\, \pi_{\omega_1,\mu_1}(L)\;.
\end{equation}
In application to the given configuration, the definition of the 3-j symbols $\Cl$ and $\Cl^*$ introduced by (\ref{X01-2},\ref{X02-2}) looks as follows:
\begin{equation}\label{Cl01}
\begin{array}{l}
\ds \pi_{\omega_1,\frac{\omega}{\omega_2}}(L) \stackrel{.}{\otimes}
\pi_{\omega_2,\frac{\omega_1}{\omega}}(L) \; \Cl \;=\;
\Cl \; \pi_{\omega,\frac{\omega_1}{\omega_2}}(L)\;,\\
\\
\ds \langle \omega_1;\frac{\omega}{\omega_2},\xi_1 | \otimes \langle \omega_2;\frac{\omega_1}{\omega},\xi_2 | \Cl | \omega, \frac{\omega_1}{\omega_2},\xi\rangle \;=\; \frac{\delta(\xi_1,\xi)\delta(\xi_2,\xi)}{S(\xi)^2}\;,
\end{array}
\end{equation}
and
\begin{equation}\label{Cl02}
\begin{array}{l}
\ds \Cl^{*}\; \pi_{\omega_2,q\frac{\omega}{\omega_1}}(L) \stackrel{.}{\otimes}
\pi_{\omega_1,q\frac{\omega_2}{\omega}}(L)\;=\;
\pi_{\omega,\frac{\omega_2}{\omega_1}}(L)\; \Cl^*\;,\\
\\
\ds \langle \omega; q\frac{\omega_2}{\omega_1},\xi | \Cl^* |\omega_2;q\frac{\omega}{\omega_1},\xi_1 \rangle \otimes | \omega_1; q\frac{\omega_2}{\omega},\xi_2\rangle \;=\;
\frac{\delta(\xi,\xi_1)\delta(\xi,\xi_2)}{S(\xi)^2}\;.
\end{array}
\end{equation}
Applying the definition (\ref{Vpi}) of the operator $\hat{V}$, as well as the relations (\ref{Cl01} and \ref{Cl01}), one obtains immediately the solution of (\ref{SLL}) in the operator form:
\begin{equation}\label{S-ans1}
\check{S}\;=\;\sum_{\omega\in\Omega}
\hat{V}_{\mu_1,\frac{\omega}{\omega_2}}\otimes
\hat{V}_{\mu_2,\frac{\omega_1}{\omega}}\;
\Cl\; \hat{V}_{\frac{\omega_1}{\omega_2},q\frac{\omega_2}{\omega_1}}\;
\Cl^*\;
\hat{V}_{q\frac{\omega}{\omega_1},\mu_2} \otimes
\hat{V}_{q\frac{\omega_2}{\omega},\mu_1}\;.
\end{equation}
The same expression in the matrix elements becomes
\begin{equation}
\begin{array}{l}
\ds \langle \xi_1^{},\xi_2^{} | \check{R} | \xi_1',\xi_2'\rangle \;=\;\\
\\
\ds =\sum_{\omega,\xi,\xi'}
V_{\mu_1,\frac{\omega}{\omega_2}}(\xi_1,\xi) 
V_{\mu_2,\frac{\omega_1}{\omega}}(\xi_2,\xi)
[\xi] 
V_{\frac{\omega_1}{\omega_2},q\frac{\omega_2}{\omega_1}}(\xi,\xi')
[\xi']
V_{q\frac{\omega}{\omega_1},\mu_2}(\xi',\xi_1')
V_{q\frac{\omega_2}{\omega},\mu_1}(\xi',\xi_2')
\end{array}
\end{equation}
This formula is written under the assumption $A=B=1$ and $S(\xi)=[\xi]$. It has the meaning of the
decomposition of the $R$-matrix with respect to irreducible representation with the help of 3-j symbols (or, the Klebsh-Gordan coefficients). Such expression, however, has a limited interest in application to the quantum integrability.

\subsection{Intertwining operator, an alternative construction}

There is another way to produce the same intertwining operator. This alternative way involves the operator $\hat{\overline{V}}$. The action of $\check{S}$ can be presented by the following diagram:
\begin{equation}
\left\{
\begin{array}{l}
\ds \pi_{\omega_1,\mu_1}\\
\otimes\\
\ds \pi_{\omega_2,\mu_2}
\end{array}\right\}
\;\;\stackrel{\hat{\overline{V}}_{z}}{\longrightarrow}\;\;
\left\{\begin{array}{lll}
\ds \pi_{\omega_1',\mu_1'} & \ds  \stackrel{\hat{V}_{\mu_1'/\mu_1''}}{\longrightarrow} & \ds \pi_{\omega_1',\mu_1''}\\
\otimes && \otimes\\
\pi_{\omega_2',\mu_2'} & \ds  \stackrel{\hat{V}_{\mu_2'/\mu_2''}}{\longrightarrow} & \ds 
\pi_{\omega_2',\mu_2''}
\end{array}
\right\}
\;\;\stackrel{\hat{\overline{V}}_{z'}}{\longrightarrow}\;\;
\left\{\begin{array}{l}
\ds \pi_{\omega_2,\mu_2}\\
\otimes\\
\ds \pi_{\omega_1,\mu_1}
\end{array}
\right\}
\end{equation}
It corresponds to the operator decomposition
\begin{equation}\label{Box}
\hat{S}\;=\;\hat{\overline{V}}_z \,\cdot\, 
\biggl( \hat{V}_{\mu_1'/\mu_1''}\, \otimes\, \hat{V}_{\mu_2'/\mu_2''}\biggr) \,\cdot\, \hat{\overline{V}}_{z'}\;,
\end{equation}
where, according to (\ref{oV3}), the arguments of the operators are given by
\begin{equation}
z\;=\;\varepsilon \, \frac{q}{\mu_1\mu_2}\, \frac{\omega_1}{\omega_2}\;,\quad
\frac{\mu_1'}{\mu_1''}\;=\;\frac{\mu_1\omega_1}{\mu_2\omega_2}\;,\quad
\frac{\mu_2'}{\mu_2''}\;=\;\frac{\mu_2\omega_1}{\mu_1\omega_2}\;,\quad
z'\;=\;\varepsilon\, \frac{\mu_1\mu_2}{q}\, \frac{\omega_1}{\omega_2}\;.
\end{equation}
The structure of these parameters can be simplified if one introduces the ``rapidities'' $x,x',y,y'$ in the same way as in (\ref{oV6}) by means of
\begin{equation}
x\;=\;\mu_1\omega_1\;,\quad x'\;=\;\frac{\omega_1}{\mu_1}\;,\quad 
y\;=\;\mu_2\omega_2\;,\quad y'\;=\;\frac{\omega_2}{\mu_2}\;.
\end{equation}
With this parametrisation,
\begin{equation}
z\;=\;\varepsilon q \frac{x'}{y} \;,\;\;
\frac{\mu_1'}{\mu_1''}\;=\; \frac{x}{y} \;,\;\;
\frac{\mu_2'}{\mu_2''}\;=\; \frac{x'}{y'} \;,\;\;
z'\;=\;\varepsilon q^{-1}  \frac{x}{y'} \;.
\end{equation}
The matrix element of $\hat{S}$ acquires the factorised ``box'' form,
\begin{equation}\label{Box2}
\langle \xi_1^{},\xi_2^{}| \hat{S} | \xi_1',\xi_2'\rangle \;=\;
\overline{V}_{\varepsilon q\frac{x'}{y}}(\xi_1^{},\xi_2^{})\;
V_{\frac{x}{y}}(\xi_1^{},\xi_1')\;
V_{\frac{x'}{y'}}(\xi_2^{},\xi_2')\;
\overline{V}_{\varepsilon q^{-1} \frac{x}{y'}}(\xi_1',\xi_2')\;.
\end{equation}
The ``box'' form is the standard form of $R$-matrix for all Ising-type models \cite{descendant,distant}.

\subsection{The ``Star-Triangle'' relation.}

In a similar to (\ref{Box}) way one can derive the Star-Triangle relation. Namely, one can consider parametrisation (\ref{oV6}) and the intertwining problem
\begin{equation}\label{ST1}
\pi_{(x,x')}\otimes \pi_{(y,y')} \; \hat{T} \;=\;
\hat{T}\; \pi_{(x/\lambda,x')}\otimes \pi_{(\lambda y,y')}\;.
\end{equation}
The operator $\hat{T}$ can be constructed in two ways,
\begin{equation}
\hat{T}_1\;=\;\biggl( \hat{V}_{\varepsilon \lambda y/ q x'} \otimes 1\biggr) \;
\hat{\overline{V}}_{\lambda}\; 
\biggl( \hat{V}_{\varepsilon q x'/y}\otimes 1\biggr)
\;\; \textrm{and}\;\;
\hat{T}_2\;=\;\hat{\overline{V}}_{\varepsilon q x'/y} \; 
\biggl(\hat{V}_{\lambda}\otimes 1\biggr) \;\hat{\overline{V}}_{\varepsilon \lambda y/ qx'}\;.
\end{equation}
The validity of both expressions can be easily verified with the help of (\ref{oV8},\ref{oV9}).
The intertwining operator $\hat{T}$ can be constructed in the third way via a 3-j symbols decomposition, but it becomes not interesting now. If the $\mathbb{V}$-form considered corresponds to an irreducible representation of the $q$-oscillator, then the intertwining operator is defined uniquely and therefore 
\begin{equation}
\hat{T}_1\;=\; R\; \hat{T}_2\;,
\end{equation}
where $R$ is a scalar constant. This equation in matrix elements becomes 
\begin{equation}\label{ST2}
\begin{array}{l}
\ds \sum_{\xi\in\mathbb{S}} \; V_x(\xi_a,\xi) \; S(\xi) 
\; \overline{V}_{xy}(\xi,\xi_c)\; V_y(\xi,\xi_b) 
\;=\;\\
\\
\ds \qquad \qquad \qquad =\;
R(x,y)\; \overline{V}_y(\xi_a,\xi_c) \; V_{xy}(\xi_a,\xi_b)
\; \overline{V}_x(\xi_b,\xi_c) \;.
\end{array}
\end{equation}
where the arguments of all operators are simplified as much as possible. Relation (\ref{ST2}) is called ``the Star-Triangle relation''. It is well known, the Yang-Baxter equation for the $R$-matrix (\ref{Box2}) follows from the Star-Triangle equation.


\section{Fock space representation}\label{F-Section}

The Fock space representation, its $\mathbb{V}$-form, a co-product of the Fock space representations and the Star-Triangle relation for the Fock space are considered in this section.

\subsection{Basic definitions}

We fix the Fock space
\begin{equation}
\mathbb{F}\;=\;\textrm{Span}\biggr\{ |a\rangle \;,\;\; a=0,1,2,\cdots\biggr\}
\end{equation}
representation in a standard way,
\begin{equation}\label{MyFock}
\begin{array}{l}
\ds 
\phi_{\omega,\lambda}(\mathcal{E}^{-})\, | a \rangle \;=\; |a-1\rangle\,\lambda\, \sqrt{1-q^{2a}}\;,
\quad
\phi_{\omega,\lambda}(\mathcal{K})\,|a\rangle \;=\; | a \rangle \, \epsilon\, q^{a+\hf}\;,\\
\\
\ds 
\phi_{\omega,\lambda}(\mathcal{K}')\,|a\rangle\;=\; | a \rangle \, \epsilon^{-1}\,q^{a+\hf}\;,\quad
\phi_{\omega,\lambda}(\mathcal{E}^{+})\, | a \rangle \;=\; | a+1 \rangle\, \lambda^{-1}\,\sqrt{1-q^{2a+2}}\;,\quad
\end{array}
\end{equation}
Here we imply the normalisation
\begin{equation}
\langle a | b \rangle \;=\; \delta_{a,b}\;,\quad 
1\;=\;\sum_{a=0}^\infty | a \rangle \, \langle a |\;.
\end{equation}
It is convenient to rewrite the Fock space representation also as 
\begin{equation}
| a \rangle \;=\; \frac{\left(\mu\mathcal{E}^{+}\right)^a}{\sqrt{(q^2;q^2)_a}} \, | 0 \rangle,\;\;
\langle a | \;=\; \langle 0 | \, \frac{\left(\mu^{-1}\mathcal{E}^{-}\right)^a}{\sqrt{(q^2;q^2)_a}}\;.
\end{equation}
The basis states here could be written as $|\omega;\lambda,a\rangle$. We will often use the shortened notation $|\lambda,a\rangle$, or even just $|a\rangle$ for them.

\subsection{$\mathbb{V}$-form of the Fock space representation.}

Consider now equation (\ref{Vdef}), defining the $\mathbb{V}$-form, in the basis (\ref{MyFock}):
\begin{equation}\label{E2-3}
\begin{array}{l}
\langle a | \phi_{\epsilon,\lambda}(\mathcal{H})(\mu) |\psi\rangle \;=\; h \langle a| \psi\rangle\;,\quad 
\langle a | \psi \rangle = \psi_a\;\;\Rightarrow\\
\\
\ds 
q^{-a-\hf} \, \biggl( \frac{\lambda}{\mu} \sqrt{1 - q^{2a+2}} \; \psi_{a+1} - \frac{\mu}{\lambda}\sqrt{1 - q^{2a}}\;  \psi_{a-1} \biggr) \;=\; h \psi_a
\end{array}
\end{equation}
This is an example of the diagonalisation problem of an unbounded operator.  It requires some comments and explanations.
\\

Here we use the following approach. We consider a truncation of the countable vector space to a finite $N+1$-dimensional one,
\begin{equation}
\mathbb{F}_{N}\;=\;\{ |0\rangle, |1\rangle, \cdots , |N\rangle\}\;.
\end{equation} 
Let then $\mathcal{H}_N$ be a reduction of the initial operator $\mathcal{H}$ to the subspace $\mathbb{F}_N$. This reduction can be constructed ambiguously since $N$-th equation in (\ref{E2-3}),
\begin{equation}
\underbrace{\frac{\lambda}{\mu} \sqrt{1-q^{2N+2}} \psi_{N+1}}_{\textrm{out of range}} - \frac{\mu}{\lambda}
\sqrt{1-q^{2N}} \psi_{N-1} = q^{N+\hf} h \psi_{N}\;,
\end{equation}
can be re-defined in different ways. Here one can see at least three ways to make a ``regularisation'' in this last equation:
\begin{equation}
\begin{array}{l}
\ds (\mathrm{I})\;:\;\; \psi_{N+1}\;=\;\frac{\mu}{\lambda} \psi_N\;,\\
\\
\ds (\mathrm{II})\;:\;\; \psi_{N+1}\;=\;-\frac{\mu}{\lambda} \psi_N\;,\\
\\
\ds (\mathrm{III})\;:\;\; \psi_{N+1}\;=\;0\;.
\end{array}
\end{equation}
The first two ways follow from the asymptotic $\ds \frac{\lambda}{\mu}\psi_{a+1}\sim\frac{\mu}{\lambda}\psi_{a-1}$, while the third way is a demonstrative example of a wrong choice.

Next, we can solve the finite-dimensional eigenvalue equation $\mathcal{H}_N|\psi\rangle \;=\; |\psi\rangle h$ for every $N$, and then look at the spectrum $\{h\}$ and at the corresponding eigenvectors in the limit $N\to 0$. If the limit exists, then we say:
\\

\emph{Given set $\{h\}$ is a branch of the generalised spectrum of the unbounded operator $\mathcal{H}$, corresponding to the type (I) (or, to type (II), etc.). For the given branch, a complete set of eigenvectors is as well provided.}
\\

It is possible that there are several different branches in a generalised spectrum (depending on a choice (I), (II), etc.), and each branch provides its own but complete set of eigenvectors.

The result of the numerical test is the following:
\begin{itemize}
\item[(I).] The limit $N\to\infty$ exists, the spectral branch is given by
\begin{equation}
h\;=\;q^{m+\hf} + q^{-m-\hf}\;,\quad m\geq 0\;.
\end{equation}
\item[(II).] The limit $N\to\infty$ exists, the spectral branch is given by
\begin{equation}
h\;=\; - (q^{m+\hf} + q^{-m-\hf})\;, \quad m\geq 0\;.
\end{equation} 
\item[(III).] Sequence of odd $N$ gives $h=\pm \ii ( q^{2m+1}-q^{-2m-1})$, $m\geq 0$. But the sequence of even $N$ gives $h=\pm\ii (q^{2m}-q^{-2m})$, $m\geq 0$. Since two subsequences have different limits, there is no limit for the choice (III).
\end{itemize}
We have a conjecture: there are only two branches for equation (\ref{E2-3}), and the branches are (I) и (II). Later we will see, the choices (I) and (II) are equivalent, and we will fix our choice on (I).

Now we can turn to the construction of the ``eigenvalues'' of equation (\ref{E2-3}). Following the convention (\ref{Vdef}), we use notations $\psi_a=\langle a | \mu,m\rangle$ with $h=\xi+\xi^{-1}$  and
\begin{equation}\label{xi-m}
\mathbb{S}\;=\;\biggl\{\xi\;=\;q^{m+\hf}\;,\quad m\geq 0\biggr\}\;.
\end{equation}
That is the set $\mathbb{S}$ for the Fock space representation.
Then,
\begin{equation}\label{PsiF}
\begin{array}{l}
\ds \langle \lambda, a | \mu, m \rangle \;=\; N_F^{-1}\, 
\left(\frac{\mu}{\lambda}\right)^a \underbrace{q^{a^2/2} \frac{\pol_a(q^{m+\hf})}{\sqrt{(q^2;q^2)_a}} q^{m(m+1)/2}}_{\chi_{a,m}}\;,\\
\\
\ds 
\langle \mu, m | \lambda, a \rangle \;=\; \left(\frac{\mu}{\lambda}\right)^{-a} 
\underbrace{q^{a^2/2+a} \frac{\pol_a(q^{m+\hf})}{\sqrt{(q^2;q^2)_a}}
\, (-)^{a} q^{m(m+1)/2}}_{\overline{\chi}_{m,a}} \;,
\end{array}
\end{equation}
where the polynomials $\pol_a(\xi)$ are defined by\footnote{Now one can mention the equivalence of the choice (I) or (II). This equivalence follows from $\pol_a(-\xi)=(-)^a \pol_a(\xi)$.}
\begin{equation}
\pol_a(\xi)\;=\;\sum_{k=0}^a \; \xi^{2k-a} \, q^{-2k(a-k)}\, \frac{(q^2;q^2)_a}{(q^2;q^2)_{k,a-k}}\;,
\end{equation}
and the reader can find a whole set of their properties in Appendix \ref{A2-Section}. The normalisation constant is  
\begin{equation}\label{NT4}
N_F \;=\; -q^{-\hf} \Theta_4\;,
\end{equation}
where the theta-constant is defined by (\ref{AT4}).
Parameter $\lambda$ for the Fock space representation and parameter $\mu$ for $\mathbb{V}$-module are shown in the left hand sides of equations (\ref{PsiF}). 
Parameter $\omega$ is common for the states and co-states there, and we omit it since the right and sides of (\ref{PsiF}) do not depend on it. What is important: the normalisation and the completeness for the states (\ref{PsiF}) are given by
\begin{equation}\label{Fnorm}
\begin{array}{l}
\ds \langle \lambda, a | \lambda, b \rangle \;=\; \sum_{m=0}^\infty \langle \lambda, a | \mu,m\rangle \, (-)^m [q^{m+\hf}]\, \langle \mu,m| \lambda, b\rangle \;=\; \delta_{a,b}\;,\\
\\
\ds 
\langle \mu,m| \mu,m'\rangle \;=\; \sum_{a=0}^\infty
\langle \mu, m|\lambda, a\rangle \langle\lambda, a | \mu, m'\rangle \;=\; \frac{(-)^m}{[q^{m+\hf}]}\; \delta_{m,m'},\quad  m,m'\geq 0\;,
\end{array}
\end{equation}
Analytical proof of these formulas follows from the properties of the polynomials $\pol_a(\xi)$
listed in the Appendix \ref{A2-Section}\footnote{We could mention that in addition to the identities 
\begin{equation}
\sum_{a=0}^\infty \overline{\chi}_{m,a}  \chi_{a,m} = (-)^m \frac{N_F}{[q^{m+\hf}]}\;,\quad
\sum_{m=0}^\infty \chi_{a,m} (-)^m [q^{m+\hf}] \overline{\chi}_{m,b} = N_F \delta_{a,b}\;,
\end{equation}
we have also
\begin{equation}
\sum_{m=-\infty}^\infty \chi_{a,m} [q^{m+\hf}] \overline{\chi}_{m,b} = 0\;,
\end{equation}
what simply follows from the antisymmetry of the summand under the change $m\to -1-m$.}.
This normalisation gives
\begin{equation}
S_m\;=\;(-)^m [q^{m+\hf}]
\end{equation}
for the relations (\ref{X5}) and (\ref{S1}), and this corresponds to 
$A=1$ in the formulas (\ref{Br},\ref{X1},\ref{X3}) and to $B=-1$ in (\ref{S1},\ref{X1-2}).
It would be helpful also to give the explicit formulas for the basis change following from (\ref{PsiF}) and from the completeness relations:
\begin{equation}\label{VF1}
\begin{array}{l}
\ds | \lambda, c \rangle \;=\; \left(\frac{\lambda}{\mu}\right)^c \; \sum_{m=0}^\infty |\mu,m\rangle \; (-)^m[q^{m+\hf}] \;\overline{\chi}_{m,c},\\
\\
\ds \langle \lambda, c | \;=\; N_F^{-1} \left(\frac{\mu}{\lambda}\right)^c \; \sum_{m=0}^\infty \chi_{c,m} \; (-)^m[q^{m+\hf}] \; \langle \mu,m|\;.
\end{array}
\end{equation}

Now we can give the definition of the Fock space representation in the terms of the definition \ref{VDEF} of the $\mathbb{V}$-form. 
\begin{definition}\label{FVdef}
A state $|f\rangle\in\mathbb{F}$ if 
\begin{itemize}
\item $\mathbb{S}:\;\xi=q^{m+\hf}$, $m\in\mathbb{Z}$, and there is the symmetry $\langle \mu, m| f \rangle = \langle \mu,-1-m|f\rangle$ (the states (\ref{PsiF}) and their normalisation have the same symmetry, but now it is taken into account by the requirement $m\geq 0$), and
\item $\langle \mu,m|f\rangle$ exponentially\footnote{This condition appears as the associativity condition
\begin{equation}
\begin{array}{l}
\ds \sum_{a=0}^\infty \langle \mu,m'|a\rangle \biggl( \sum_{m=0}^\infty \langle a | \mu,m\rangle (-)^m [q^{m+\hf}] \langle \mu,m| f\rangle\biggr)\;=\\
\\
\ds =\; \sum_{m=0}^\infty \biggl( \sum_{a=0}^\infty 
\langle \mu,m'|a\rangle \langle a | \mu,m\rangle \biggr) (-)^m [q^{m+\hf}] \langle \mu,m| f\rangle\;.
\end{array}
\end{equation}
}
decays when $m\to \infty$.
\end{itemize}
\end{definition}

\subsubsection{Co-product $\mathbb{F}\otimes\mathbb{F}$ in the terms of $\mathbb{V}$-form.}

Now we can see how the co-product works when one uses the $\mathbb{V}$-form. We consider the co-product of two Fock spaces,
\begin{equation}
\phi_{\omega_1,\lambda_1}\otimes \phi_{\omega_2,\lambda_2} (\Delta(L))\;.
\end{equation}
The $\mathbb{V}$-forms for these representations are defined by relations (\ref{PsiF}), while all subsequent steps are described in Section \ref{V-Section}.
The (non-normalised) $\mathbb{V}$-basis (\ref{X01}) for the co-product is
\begin{equation}\label{FF}
\left\{
\begin{array}{l}
\ds 
|\!|\omega; \frac{\omega_1}{\omega_2},m\rangle \;=\; 
|\omega_1;\frac{\omega}{\omega_2},m\rangle \otimes |\omega_2;\frac{\omega_1}{\omega},m\rangle\;,\\
\\
\ds 
\langle a,b |\!|\omega; \frac{\omega_1}{\omega_2},m\rangle \;=\; \left(\lambda_1^{-1}\frac{\omega}{\omega_2}\right)^a 
\left(\lambda_2^{-1}\frac{\omega_1}{\omega}\right)^b \; \chi_{a,m} \, \chi_{b,m}\;,
\end{array}\right.
\end{equation}
where the matrices $\chi$ and $\overline{\chi}$ are defined in (\ref{PsiF}), 
these states imply the representations (\ref{X3}) and its dual with $\xi=q^{m+\hf}$ and $A=1$. Evidently, the definition \ref{FVdef} of the Fock space representation can be extended to the direct product of two $\mathbb{V}$-forms, so that (\ref{FF}) is the similar Fock space representation.  Therefore, one can use (\ref{VF1}) for the reduction of the states $|\!|\omega;\frac{\omega_1}{\omega_2},m\rangle$ back to the occupation numbers basis, to $|\!|\omega,c\rangle$,
\begin{equation}
|\!|\omega,c\rangle \;=\; \left(\frac{\omega_2}{\omega_1}\right)^c\; \sum_{m=0}^\infty |\!|\omega,\frac{\omega_1}{\omega_2},m\rangle \, (-)^m [q^{m+\hf}]\, \overline{\chi}_{m,c}\;.
\end{equation}
It gives
\begin{equation}
\langle a,b |\!| \omega,c\rangle \;=\; 
\left(\lambda_1^{-1}\frac{\omega}{\omega_2}\right)^a 
\left(\lambda_2^{-1}\frac{\omega_1}{\omega}\right)^b 
\left(\frac{\omega_2}{\omega_1}\right)^c\;
\left(
M_F^{-1} \sum_{m=0}^\infty \chi_{a,m}\chi_{b,m}\overline{\chi}_{m,c} (-)^m[q^{m+\hf}]\right) \;,
\end{equation}
where new proper normalisation $M_F$ is introduced (see (\ref{AFM} for the details),
\begin{equation}\label{F-M}
M_F\;=\;-q^{-\hf} (q;q)_\infty\;.
\end{equation}
A remarkable object has been arisen,
\begin{equation}
M_F^{-1} \sum_{m=0}^\infty \chi_{a,m}\chi_{b,m}\overline{\chi}_{m,c}(-)^m[q^{m+\hf}]\;=\;
\frac{(-q)^c}{\sqrt{(q^2;q^2)_{a,b,c}}} \{a,b,c\}\;,
\end{equation}
where
\begin{equation}\label{abc-P}
\{a,b,c\} = M_F^{-1}
q^{(a^2+b^2+c^2)/2} \sum_{m=0}^\infty (-)^m q^{3m(m+1)/2} [q^{m+\hf}] P_a(q^{m+\hf})
P_b(q^{m+\hf})P_c(q^{m+\hf})
\end{equation}
The following identity has been verified numerically:
\begin{equation}\label{abc}
\{a,b,c\} = (q^2;q^2)_{a,b,c} \,
q^{ab+ac+bc}\, \sum_{k=0}^{\min(a,b,c)} (-)^k \frac{q^{3k^2-k-2k(a+b+c)}}{(q^2;q^2)_{k,a-k,b-k,c-k}}\;.
\end{equation}
Thus, one obtains the expression for the Klebsh-Gordan coefficients (or, 3-j symbols) in the initial Fock space basis:
\begin{equation}
\langle a, b |\!|\omega,c\rangle\;=\;
\left(\lambda_1^{-1}\frac{\omega}{\omega_2}\right)^a 
\left(\lambda_2^{-1}\frac{\omega_1}{\omega}\right)^b
\left(-q\frac{\omega_2}{\omega_1}\right)^c \, \frac{\{a,b,c\}}{\sqrt{(q^2;q^2)_{a,b,c}}}\;.
\end{equation}

The left Fock space co-module could be obtained in the same way as it is described in Section \ref{V-Section}, but with a tiny modification. Our normalisation (\ref{Fnorm}) implies $B=-1$ for the co-module (\ref{X1-2}). ``Wrong'' sign can be compensated by the change of the signs of $\omega$ and $\mu$, so that the formula (\ref{X02}) in our current situation takes the form 
\begin{equation}
\langle \omega_1; q\frac{\omega}{\omega_2},m | \otimes 
\langle \omega_2; q\frac{\omega_1}{\omega},m|
\;=\;
\langle -\omega; -q\frac{\omega_1}{\omega_2},m |\!|\;.
\end{equation}
Changing the sign in front of $\omega$ in both hand sides, one comes to the acceptable form of the co-product for the co-module,
\begin{equation}
\left\{
\begin{array}{l}
\ds \langle \omega; -q\frac{\omega_1}{\omega_2},m |\!| \;=\;
\langle \omega_1; -q\frac{\omega}{\omega_2},m | \otimes 
\langle \omega_2; -q\frac{\omega_1}{\omega},m|\;,\\
\\
\ds
\langle \omega; -q\frac{\omega_1}{\omega_2},m|\!| a,b\rangle \;=\;
\left(-q\lambda_1^{-1}\frac{\omega}{\omega_2}\right)^{-a} 
\left(-q\lambda_2^{-1}\frac{\omega_1}{\omega}\right)^{-b} 
\; \overline{\chi}_{m,a} \, \overline{\chi}_{m,b}\;,
\end{array}\right.
\end{equation}
what gives
\begin{equation}\label{FF2}
\begin{array}{l}
\ds \langle \omega, c|\!| a,b \rangle \;=\;
\left(-q\lambda_1^{-1}\frac{\omega}{\omega_2}\right)^{-a} 
\left(-q\lambda_2^{-1}\frac{\omega_1}{\omega}\right)^{-b} 
\left(-q\frac{\omega_1}{\omega_2}\right)^c\;\times\\
\\
\ds \qquad \qquad \times
\underbrace{\biggl( M_F^{-1}\sum_{m=0}^\infty \chi_{c,m} (-)^m [q^{m+\hf}] \overline{\chi}_{m,a} \overline{\chi}_{m,b} \biggr)}_{\ds \frac{(-q)^{a+b}}{\sqrt{(q^2;q^2)_{a,b,c}}}\{a,b,c\}}
\end{array}
\end{equation}
Thus, the final formulas for the Klebsh-Gordan coefficients become completely symmetric,
\begin{equation}\label{FKG}
\left\{
\begin{array}{l}
\ds 
\langle a,b |\!| \omega,c\rangle \;=\; 
\left(\lambda_1^{-1}\frac{\omega}{\omega_2}\right)^a 
\left(\lambda_2^{-1}\frac{\omega_1}{\omega}\right)^b
\left(-q\frac{\omega_2}{\omega_1}\right)^c\;
\frac{\{a,b,c\}}{\sqrt{(q^2;q^2)_{a,b,c}}}\;,\\
\\
\ds 
\langle \omega,c|\!|a,b\rangle \;=\; 
\left(\lambda_1^{-1}\frac{\omega}{\omega_2}\right)^{-a} 
\left(\lambda_2^{-1}\frac{\omega_1}{\omega}\right)^{-b}
\left(-q\frac{\omega_1}{\omega_2}\right)^c\;
\frac{\{a,b,c\}}{\sqrt{(q^2;q^2)_{a,b,c}}}\;,
\end{array}\right.
\end{equation}
as it should be for the symmetric initial representation (\ref{MyFock}).
The co-product states are normalised (details of this calculation can. be found 
in (\ref{AFnorm})):
\begin{equation}\label{FKGn}
\ds \langle \omega,c|\!| \omega',c'\rangle \;=\; \frac{2\pi}{(q^2;q^2)_\infty} \delta(\varphi-\varphi')\delta_{c,c'}\;,\quad \omega=\EXP^{\ii\varphi},\;\;\omega'=\EXP^{\ii\varphi'}.
\end{equation}
Thus, using the method of $\mathbb{V}$-forms, one obtains the Klebsh-Gordan coefficients (\ref{FKG}), and their normalisation provides 
\begin{equation}
\mathbb{F}\otimes\mathbb{F}=\mathbb{T}\otimes\mathbb{F}\;,\quad
\textrm{where}\quad \mathbb{T}\,\ni\,\omega\;.
\end{equation}

\subsection{Alternative derivation of the Klebsh-Gordan coefficients.}

The expressions (\ref{FKG}) can be obtained in a straightforward way.
First, one can construct the vacuum and co-vacuum for a fixed value of $\omega$. The vacuum equations
\begin{equation}
\Delta(\mathcal{E}^{-}) |\!|\omega,0 \rangle\;=\;0\;,\quad
\Delta(\mathcal{K}) |\!|\omega,0\rangle \;=\; |\!|\omega,0\rangle\,\omega\,q^{\hf}
\end{equation}
have the unique solution 
\begin{equation}\label{vac1}
\langle a,b |\!| \omega,0 \rangle \;=\; 
\left(\lambda_1^{-1}\frac{\omega}{\omega_2}\right)^a
\left(\lambda_2^{-1}\frac{\omega_1}{\omega}\right)^b
\;
\frac{q^{ab}}{\sqrt{(q^2;q^2)_{a,b}}}\;.
\end{equation}
Next,
\begin{equation}\label{fep}
\Delta(\mathcal{E}^{+})\; |\!|\omega,c\rangle\;=\;
|\!|\omega,c+1\rangle \, \sqrt{1-q^{2c+2}}
\end{equation}
gives the formula (\ref{FKG}), in which the coefficients $\{a,b,c\}$ are defined by the recursion
\begin{equation}
\{a,b,c+1\}\;=\;q^{a+b}\{a,b,c\}-q^{-1}(1-q^{2a})(1-a^{2b})\{a-1,b-1,c\}\;,\quad
\{a,b,0\}\;=\;q^{ab}\;,
\end{equation}
and the solution of this recursion is given by the formula (\ref{abc}). 
The dual Klebsh-Gordan coefficients can be obtained from the similar equations,
\begin{equation}\label{fem}
\langle \omega,0|\!|\, \Delta(\mathcal{E}^{+}) = 0,\;\;
\langle \omega,0|\!|\,\Delta(\mathcal{K}) = \omega\, q^{\hf} 
\langle \omega,0|\!|,\;\;
\langle \omega,c|\!|\,\Delta(\mathcal{E}^{-})=\sqrt{1-q^{2c+2}} \langle \omega,c+1|\!|.
\end{equation}

\subsection{Similarity operators $\hat{V}$ and $\hat{\overline{V}}$ and the basic integrable model.}

The similarity $\hat{V}$-type operator for the initial Fock space representation is primitive,
\begin{equation}
\hat{V}_{\lambda,\lambda'} \; \phi_{\omega,\lambda'} \;=\;
\phi_{\omega,\lambda} \; \hat{V}_{\lambda,\lambda'}\quad
\Rightarrow\quad 
\langle \lambda,a| \hat{V}_{\lambda,\lambda'} | \lambda',a'\rangle \;=\; \left(\frac{\lambda'}{\lambda}\right)^a \;\delta_{a,a'}\;,
\end{equation}
and therefore it is not interesting.

But the similarity operator in $\mathbb{V}$-basis 
\begin{equation}
\langle m | \hat{V}_{\mu,\mu'} | m' \rangle \;=\;
\sum_{a=0}^\infty \langle \mu,m | \lambda,a\rangle \langle \lambda,a | \mu',m'\rangle 
\;\stackrel{def}{=}\; 
V_{\mu/\mu'}(m,m')\;,
\end{equation}
is given by (see eq. (\ref{A7}))
\begin{equation}\label{F-V-1}
\begin{array}{l}
\ds 
V_{\mu/\mu'}(m,m')\;=\;
N_F^{-1} q^{m(m+1)/2+m'(m'+1)/2}\;\times\\
\\
\ds \qquad \times\;
\frac{\ds (
\frac{\mu'}{\mu} q^{2+m-m'}, \frac{\mu'}{\mu} q^{2-m+m'},
\frac{\mu'}{\mu} q^{3+m+m'},\frac{\mu'}{\mu} q^{1-m-m'};q^2)_\infty}
{\ds (q^2\frac{\mu^{\prime 2}}{\mu^2};q^2)_\infty}\;.
\end{array}
\end{equation}
Let us recall the properties of this operator (now the properties can be verified analytically). First, the symmetry,
\begin{equation}
V_{\mu/\mu'}(m,m')\;=\;
V_{\mu/\mu'}(m',m)\;=\;
V_{\mu/\mu'}(-1-m,m')\;=\;\cdots
\;,
\end{equation}
then, the normalisation,
\begin{equation}
V_{\mu/\mu}(m,m')\;=\;\frac{(-)^m}{[q^{m+\hf}]}\;\delta_{m,m'}\;,\quad 
m,m'\;\geq \; 0\;,
\end{equation} 
and the summation (transitivity) formula,
\begin{equation}\label{F-sum}
\sum_{m=0}^\infty 
V_{\mu/\mu'}(m,m') \, (-)^{m'} [q^{m'+\hf}]\,
V_{\mu'/\mu''}(m,m'') \;=\;
V_{\mu/\mu''}(m,m'')\;.
\end{equation}
There is one more property,
\begin{equation}\label{myinv}
V_z(m,m') V_{q^2/z}(m,m')\;=\;N_F^{-2}\, \frac{(z,q^2/z;q)_\infty}{(-z/q,-q/z;q)_\infty}\;.
\end{equation}
Now one can turn to the operator $\hat{\overline{V}}_z$ defined by the relations (\ref{oV1},\ref{oV8},\ref{oV4}). It turns out, we must choose $\varepsilon=-1$ and define 
\begin{equation}\label{F-V-2}
\overline{V}_z(m,m')\;=\;V_{-q/z}(m,m')\;.
\end{equation}
As the result, the Star-Triangle relation for the $\mathbb{V}$-form of the Fock space representation holds:
\begin{equation}\label{mystst}
\begin{array}{l}
\ds \sum_{m=0}^\infty\;
\; V_{x}(m_a,m) \; S_m\; V_{-q/xy}(m,m_c)\; V_{y}(m_b,m) \; \;=\;\\
\\
\ds \qquad\qquad =\;
R_{x,y,-q/xy}\; V_{-q/y}(m_a,m_c) \; V_{xy}(m_a,m_b)\; V_{-q/x}(m_b,m_c) \; \;,
\end{array}
\end{equation}
where the scalar factor (to be calculated separately) is 
\begin{equation}
R_{x,y,-q/xy}\;=\; N_F\; \kappa_x \, \kappa_y \, \kappa_{-q/xy}\;,\quad
\kappa_x\;=\;\frac{(x;q)_\infty}{(-q/x;q)_\infty}\;.
\end{equation}
The physical regime for this model is
\begin{equation}
0<-q<x<1
\end{equation}
The standard arguments relating partition functions to the parameters of the star-triangle relation (\ref{mystst}) and inversion relation (\ref{myinv}) allows one to deduce 
\begin{equation}
\langle S \rangle \;=\; N_F\;,\quad
\langle V_x\rangle \;=\; \frac{\kappa_x}{N_F} \, z_x
\end{equation}
where $\langle S\rangle$ is the partition function per site $S_m$ and $\langle V_x\rangle$ is the partition function per edge $V_x(m,m')$ of the lattice in the thermodynamic limit, and
\begin{equation}
\log z_x\;=\;\sum_{n=1}^\infty \frac{x^n - (q^2/x)^n}{n(1-q^n)(1+(-q)^n)}\;.
\end{equation}
This expression literally coincide with the partition function for the ``distant descendant of the six-vertex model'' \cite{distant}.

\subsection{The Fock space intertwining operators}

This section seems to be incomplete without various explicit forms of the intertwining operator 
$\hat{S}$.

Let the Fock space representations are defined in their $\mathbb{V}$-forms.
Then, considering the intertwining relation (\ref{SLL}),
\begin{equation}\label{SLL-2}
\pi_{\omega_1,\mu_1}(L) \, \stackrel{.}{\otimes}\, \pi_{\omega_2,\mu_2}(L)
\;\check{S}
\;=\;
\check{S}\; 
\pi_{\omega_2,\mu_2}(L) \, \stackrel{.}{\otimes}\, \pi_{\omega_1,\mu_1}(L)\;.
\end{equation}
one could repeat the reasoning of the definitions (\ref{Cl01},\ref{Cl02}), but also the relations (\ref{FF},\ref{FF2}) could be taken into account, so that (\ref{Cl01},\ref{Cl02}) could be modified,
\begin{equation}\label{Cl012}
\begin{array}{l}
\ds \pi_{\omega_1,\frac{\omega}{\omega_2}}(L) \stackrel{.}{\otimes}
\pi_{\omega_2,\frac{\omega_1}{\omega}}(L) \; \Cl \;=\;
\Cl \; \pi_{\omega,\frac{\omega_1}{\omega_2}}(L)\;,\\
\\
\ds \langle \omega_1;\frac{\omega}{\omega_2},m_1 | \otimes \langle \omega_2;\frac{\omega_1}{\omega},m_2 | \Cl | \omega, \frac{\omega_1}{\omega_2},m \rangle \;=\; \frac{\delta_{m_1,m}\delta_{m_2,m}}{S_m^2}\;,
\end{array}
\end{equation}
and
\begin{equation}\label{Cl022}
\begin{array}{l}
\ds \Cl^{*}\; \pi_{\omega_2,-q\frac{\omega}{\omega_1}}(L) \stackrel{.}{\otimes}
\pi_{\omega_1,-q\frac{\omega_2}{\omega}}(L)\;=\;
\pi_{\omega,-\frac{\omega_2}{\omega_1}}(L)\; \Cl^*\;,\\
\\
\ds \langle \omega; -q\frac{\omega_2}{\omega_1},m | \Cl^* |\omega_2;-q\frac{\omega}{\omega_1},m_1 \rangle \otimes | \omega_1; -q\frac{\omega_2}{\omega},m_2\rangle \;=\;
\frac{\delta_{m,m_1}\delta_{m,m_2}}{S_m^2}\;,
\end{array}
\end{equation}
so that the answer (\ref{S-ans1}) could be modified slightly,
\begin{equation}
\hat{S}\;=\;\hat{V}_{\mu_1,\frac{\omega}{\omega_1}}\otimes \hat{V}_{\mu_2,\frac{\omega_1}{\omega}}\; \Cl\; 
\hat{V}_{\frac{\omega_1}{\omega_2},-q\frac{\omega_2}{\omega_1}}\;
\Cl^*\; 
\hat{V}_{-q\frac{\omega}{\omega_1},\mu_2}\otimes \hat{V}_{-q\frac{\omega_2}{\omega},\mu_1}\;.
\end{equation}
The same intertwining operator in the same basis (up to a scalar multiplier) is given by the ``box'' expression (\ref{Box}) with the operators $\hat{V}$ and $\hat{\overline{V}}$ given by  (\ref{F-V-1}) and (\ref{F-V-2}) with the same parameters as given after the equation (\ref{Box}).

From the other hand side, if the $q$-oscillators are initially defined in the ``old-fashion'' representation (\ref{MyFock}), then the solution for $\hat{S}$ is more simple. According to our convention, now we try to intertwine 
\begin{equation}\label{SLL-3}
\phi_{\omega_1,\lambda_1}(L) \, \stackrel{.}{\otimes}\, \phi_{\omega_2,\lambda_2}(L)
\;\check{S}
\;=\;
\check{S}\; 
\phi_{\omega_2,\lambda_2}(L) \, \stackrel{.}{\otimes}\, \phi_{\omega_1,\lambda_1}(L)\;.
\end{equation}
Since we intertwine the canonical representations (\ref{MyFock}), the canonical Klebsh-Gordan coefficients (\ref{FKG}) are to be taken; and since the operator $\hat{S}$ in the basis $|\!|\omega,c\rangle$ is just the identity operator, then 
\begin{equation}
\begin{array}{l}
\ds \langle a,b | \check{S} | a',b'\rangle \;=\\
\\
\ds =\; \frac{1}{2\pi}\sum_{c=0}^\infty\int d \varphi \sum_c \langle \omega_1;\lambda_1,a|\otimes\langle \omega_2;\lambda_2,b|\!| \omega,c\rangle 
\langle \omega,c |\!| \omega_2;\lambda_2,a'\rangle \otimes |\omega_1;\lambda_1,b'\rangle
\end{array}
\end{equation}
where all ingredients are defined by (\ref{FKG}). The integral over $\varphi$ can be taken, it leads to a simple delta-function. The sum over $c$ gives an infinite series in
$\omega_1/\omega_2$.


\section{Representation $\mathbb{V}_\gamma$}\label{F2-Section}

In this section we consider a natural generalisation of the Fock-space $\mathbb{V}$-form.
This generalisation provides in particular the rational Kashiwara-Miwa model.

\subsection{The generalisation}

We consider now the representation (\ref{Hom2}) with 
\begin{equation}\label{w2-1}
\uop | m \rangle \;=\; | m \rangle \, \gamma q^m\;,\quad
\vop | m \rangle \;=\; | m+1 \rangle\;,\quad
\langle m | m' \rangle \;=\; \frac{\delta_{m,m'}}{[\gamma q^m ]}\;.
\end{equation}
It implies 
\begin{equation}\label{Vrep}
\left\{
\begin{array}{l}
\ds \pi_{\omega,\mu}^{(\gamma)}(\mathcal{K}_0)\, |m \rangle \;=\; \biggl( |m+1 \rangle - | m-1 \rangle \biggr) / [\gamma q^m]\;,\\
\\
\ds \pi_{\omega,\mu}^{(\gamma)}(\mathcal{E}^{+})\, |  m \rangle \;=\; \mu^{-1}\, \biggl( | m-1 \rangle \gamma q^m  - | m+1 \rangle \gamma^{-1} q^{-m} \biggr) / [\gamma q^m]\;,\\
\\
\ds \pi_{\omega,\mu}^{(\gamma)}(\mathcal{E}^{-})\, | m \rangle \;=\; \mu\, \biggl( | m+1 \rangle \gamma q^m  - | m-1 \rangle \gamma^{-1} q^{-m} \biggr) / [\gamma q^m]\;.
\end{array}\right.
\end{equation}
This representation corresponds to (\ref{X1}) with 
\begin{equation}
\mathbb{S} \;=\; \biggl\{\gamma q^m,\;m\in\mathbb{Z}\biggr\}\;.
\end{equation}
As before, we will denote the states as 
\begin{equation}
| m \rangle \;\to\; |\mu,m\rangle\;,\quad \textrm{or}\quad |m\rangle \;\to\; |\omega;\mu,m\rangle\;,
\end{equation}
when the parameters $\omega$ and $\mu$ will become important. Parameter $\gamma$ will be always implied.

For the complete definition of the representation $\pi_{\omega,\mu}^{(\gamma)}$ we need to impose the extra requirement of the exponential decay of the matrix elements $\langle m | \psi\rangle $, $|\psi\rangle \in \mathbb{V}$.

\subsection{Decomposition $\mathbb{V}_\gamma=\mathbb{F}\oplus\mathbb{F}$.}

There is the 
\begin{proposition}
The representation $\pi_{\omega,\mu}^{(\gamma)}$ is a direct sum of two Fock-space representations.
\end{proposition}
Corresponding matrix elements are given by
\begin{equation}\label{F2V}
\begin{array}{l}
\ds 
\langle \epsilon,a |\mu, m \rangle \;=\; N_\gamma^{-1}\, \mu^a\,
\underbrace{\epsilon^{a+m} \,  \gamma^m \, q^{a^2/2+m^2/2} \, \frac{\pol_a(\gamma q^m)}{\sqrt{(q^2;q^2)_a}}}_{\chi_{(\epsilon,a),m}}\;,\\
\\
\ds \langle \mu, m | \epsilon,a \rangle \;=\;  \mu^{-a}\, 
\underbrace{(-\epsilon)^{a+m} \,  \gamma^m\, q^{a^2/2 +a + m^2/2}\, 
\frac{\pol_a(\gamma q^m)}{\sqrt{(q^2;q^2)_a}}}_{\overline{\chi}_{m,(\epsilon,a)}}\;.
\end{array}
\end{equation}
Here $\epsilon=\pm 1$ is the sign, and the normalisation scalar is
\begin{equation}
N_\gamma\;=\;-2\BH(\gamma)\;,
\end{equation}
where $\BH$ is defined by (\ref{ATTH}). Comparing this normalisation factor with (\ref{NT4}),
one sees the extra multiplier $2$ -- this is because now $m\in\mathbb{Z}$.

The Fock states in (\ref{F2V}) satisfy the relations (\ref{MyFock}) with small variations,
\begin{equation}
\mathcal{K}_0 |\epsilon, a\rangle = |\epsilon,a\rangle \epsilon q^{a+\hf} \;,\quad
|\epsilon, a\rangle \;=\; \frac{(\mathcal{E}^{+})^a}{\sqrt{(q^2;q^2)_a}}\, 
|\epsilon, 0\rangle\;,
\end{equation}
what means that $\epsilon$ is the parity, and $\lambda$-parameter just equals to $1$. The completeness relations now become
\begin{equation}
\begin{array}{l}
\ds 
\sum_{a=0}^\infty \sum_{\epsilon=\pm} \langle \mu, m | \epsilon, a \rangle \, \langle \epsilon, a | \mu, m' \rangle \;=\;
\frac{\delta_{m,m'}}{[\gamma q^m]}\;,\\
\\
\ds 
\sum_{n\in\mathbb{Z}} \langle \epsilon, a | \lambda, m \rangle\, [\gamma q^m]\,  \langle \mu, m | \epsilon',a'\rangle\;=\;
\delta_{\epsilon,\epsilon'} \delta_{a,a'}\;.
\end{array}
\end{equation}
The previous Fock space $\mathbb{V}$-form is related to the newly defined one by the requirement $\gamma=q^{\hf}$ and 
\begin{equation}
\textrm{Old state}\;\; |\mu,m\rangle\;=\;\textrm{Newly defined}\;\;\frac{1}{2} \biggl( |\mu,m\rangle + |\mu,-1-m\rangle\biggr)\;.
\end{equation}
For the old Fock space only symmetric $|+,a\rangle$ survives, while the anti-symmetric $|-,a\rangle$ identically equals to zero.

\subsection{Intertwining operator $\hat{V}$ and projection decomposition.}

Let $\mathbb{V}^{(\epsilon)}$ be the component of $\mathbb{V}$, corresponding to the $\epsilon$-subspace. Let further
\begin{equation}
V_{\mu/\mu'}^{(\epsilon)}(m,m')\;=\;\sum_{a=0}^\infty \, \langle \mu, m | \epsilon,a\rangle\, \langle \epsilon, a| \mu',m\rangle\;.
\end{equation}
Using (\ref{A7}), one obtains 
\begin{equation}\label{V-V-1}
\begin{array}{l}
\ds V_x^{(\epsilon)}(m,m')\;=\; N_\gamma^{-1}\, (-\epsilon)^m \epsilon^{m'} \, q^{m^2/2+m^{\prime 2}/2} \, \gamma^{m+m'}\;\times\\
\\
\ds \times\;
\frac{(x^{-1}\gamma^2 q^{2+m+m'},x^{-1}\gamma^{-2}q^{2-m-m'},x^{-1}q^{2+m-m'},x^{-1}q^{2-m+m'};q^2)_\infty}{(q^2/x^2;q^2)_\infty}\;.
\end{array}
\end{equation}
The summation formula now acquires the orthogonality structure, 
\begin{equation}\label{V-sum}
\sum_{m'\in\mathbb{Z}} V_x^{(\epsilon)}(m,m') \, [\gamma q^m]\, V_y^{(\epsilon')}(m',m'')\;=\;
\delta_{\epsilon,\epsilon'} V_{xy}^{(\epsilon)}(m,m'')\;.
\end{equation}
The normalisation of $\hat{V}^{(\epsilon)}$ gives the projection matrices to $\mathbb{V}^{(\epsilon)}$,
\begin{equation}
[\gamma q^m]\, V_1^{(\epsilon)}(m,m') = \mathcal{P}_{m,m'}^{(\epsilon)},\;\;
\mathcal{P}^{(\epsilon)}\mathcal{P}^{(\epsilon')} = \delta_{\epsilon,\epsilon'}\mathcal{P}^{(\epsilon)},\;\;
\mathcal{P}^{(+)}+\mathcal{P}^{(-)} = 1.
\end{equation}

\subsection{Co-product $\mathbb{V}_\gamma\otimes\mathbb{V}_\gamma$.}

Since $\mathbb{V}_\gamma\;=\;\mathbb{F}\oplus\mathbb{F}$, its direct product has nothing special. Only one thing to be mentioned, is the current version of the formula (\ref{abc-P}):
\begin{equation}
\{a,b,c\}\;=\;M_\gamma^{-1}\, q^{a^2+b^2+c^2)/2}\,
\sum_{m\in\mathbb{Z}} (-)^m q^{3m^2/2}\gamma^{3m} [\gamma q^m]
\pol_a(\gamma q^m)\pol_b(\gamma q^m)\pol_c(\gamma q^m),
\end{equation}
where $\gamma$-independent $\{a,b,c\}$ is the same as in (\ref{abc}), and the pre-factor is given by
\begin{equation}
M_\gamma\;=\;
\sum_{m\in\mathbb{Z}} (-)^m q^{3m^2/2}\gamma^{3m}[\gamma q^m]=-\frac{\BH(\gamma) \theta_3(\gamma)}{(q^2;q^2)_\infty}\;,
\end{equation}
where the theta-functions are defined in (\ref{ATTH}).

\subsection{The Star-Triangle relation and the integrable rational Kashiwara-Miwa model.}

As we observed, the representation $\pi_{\omega,\mu}^{(\gamma)}$ with generic $\gamma$ is reducible, so that there are no arguments in favour of a uniqueness of the Star-Triangle map (\ref{ST1})\footnote{I spend essential amount of efforts for intensive numerical tests looking for the Star-Triangle and Star-Star relations with generic $\gamma$. Result is negative.}. However, there are some selected values of $\gamma$ (in addition to $\gamma=q^{\hf}$ from the previous section) such that the other irreducible Fock space representations arise. These additional irreducible Fock space representations are not the representations of the generalised algebra 
$\mathcal{O}_q[\mathcal{K},\mathcal{K}']$, but they are the irreducible representations of its subalgebra $\mathcal{O}_q\subset \mathcal{O}_q[\mathcal{K},\mathcal{K}']$. The selected values of $\gamma$ are 
\begin{equation}\label{vyd}
\gamma\;=\;\ii\, q^{\nu}\;,\quad \nu\in\mathbb{Z}/2\;.
\end{equation}
The set of such values of $\gamma$ factorises into two equivalence classes,
\begin{equation}
\nu\;=\;0\quad\textrm{or}\quad \nu\;=\; \hf.
\end{equation}
The state space can be divided into even and odd subspaces.
\begin{equation}
\mathbb{V}_s\;=\;\biggl\{ \psi_m\;:\;\; \psi_m=\psi_{2\nu-m}\biggr\}\;,\quad
\mathbb{V}_a\;=\;\biggl\{ \psi_m\;:\;\; \psi_m=-\psi_{2\nu-m}\biggr\}\;.
\end{equation}
The direct verification shows that these subspaces are the invariant subspaces for $\mathcal{E}^{\pm}$, and therefore they are irreducible representations of $\mathcal{O}_q$. However, the operator $\mathcal{K}_0$ changes the parity, $\mathbb{V}_s\to\mathbb{V}_a$, therefore we do not talk about $\mathcal{O}_q[\mathcal{K},\mathcal{K}']$.

It is convenient to give more simple form of the weight $V$. To do it, we note that 
\begin{equation}
V_x^{(+)}(m,m')+V_x^{(-)}(m,m')\;=\;0\;\;\textrm{if}\;\;
m+m'=1\mod 2\;.
\end{equation}
The case when both $m$ and $m'$ are odd is equivalent to the case when both $m$ and $m'$ are even but $\nu$ is shifted by $2$. Therefore, we can restrict ourselves to the case of even $m$ and $m'$.

Thus, we could define 
\begin{equation}\label{VKM}
\Vop_{\!x}(m,m')\;=\; \frac{V^{(\pm)}_x(2m,2m')}{V^{(\pm)}_z(0,0)}\;=\;
\left(\frac{q}{x}\right)^{2m} \, \frac{(x;q^2)_{m-m'}}{(q^2/x;q^2)_{m-m'}}\,
\frac{(\gamma^2 x;q^2)_{m+m'}}{(q^2/\gamma^2 x;q^2)_{m+m'}}\;.
\end{equation}
In application to the integrable models let also 
\begin{equation}
\boldsymbol{S}_m\;=\;\frac{[\gamma q^{2m}]}{[\gamma]} \;,
\end{equation}
and
\begin{equation}
\Phi(x)\;=\;\frac{1}{2 [\gamma] V^{(\pm)}_z(0,0)}\;=\;
\frac{(q^2,q^2/x^2,q^2\gamma^2,q^2/\gamma^2;q^2)_\infty}{(q^2/x,q^2/x,q^2\gamma^2/x,q^2/\gamma^2x;q^2)_\infty}\;.
\end{equation}
The second weight, $\overline{V}$, can be defined now by 
\begin{equation}
\overline{\Vop}_x(a,b)\;=\;\Vop_{q/x}(a,b)\;.
\end{equation}
The symmetry properties of the weight (\ref{VKM}) are the following:
\begin{equation}
\begin{array}{ll}
\ds \Vop_x(a,b) \;=\; \Vop_x(b,a)\;, & \ds \Vop_1(a,b)\;=\;\frac{\delta_{a,b}}{\boldsymbol{S}_a}\;,\\
\\
\ds \Vop_x(a,b)\,\Vop_{q^2/x}(a,b)\;=\;1\;, & \ds
\Vop_{q}(a,b)\;=\;1\;.
\end{array}
\end{equation}
Also, the summation formula for the weight (\ref{VKM}) now becomes
\begin{equation}
\sum_{b\in\mathbb{Z}} \Vop_x(a,b) \, \boldsymbol{S}_b \, \Vop_y(b,c)\;=\;
\frac{\Phi(x)\Phi(y)}{\Phi(xy)}\, \Vop_{xy}(a,c)\;.
\end{equation}
The inversion relation is the particular case of the summation formula,
\begin{equation}
\sum_{b\in\mathbb{Z}} \Vop_x(a,b) \, \boldsymbol{S}_b \, \Vop_{1/x}(b,c)\;=\;
\Phi(x)\Phi(1/x)\, \frac{\delta_{a,c}}{\boldsymbol{S}_a}\;.
\end{equation}
All relations so far are valid for generic $\gamma$. But the Star-Triangle relation,
\begin{equation}
\begin{array}{l}
\ds \sum_{d\in\mathbb{Z}} \, \boldsymbol{S}_d \, \Vop_x(a,d) \, \Vop_y(b,d) \,\Vop_z(c,d)\;=\\
\\
\ds \qquad =\; 
\frac{\kappa(x)\kappa(y)\kappa(z)}{\kappa(1)}\, 
\Vop_{q/x}(b,c) \, \Vop_{q/y}(a,c) \Vop_{q/z}(a,b)\;,\quad xyz\;=\;q\;,
\end{array}
\end{equation}
is valid only for the selected values (\ref{vyd}) of $\gamma$. The scalar functions $\kappa$ here are: 
\begin{equation}
\begin{array}{l}
\ds \gamma\;=\;\ii\;,\qquad \kappa(x)\;=\frac{(q^2x^2;q^4)_\infty}{(q^4/x^2;q^4)_\infty}\;,\\
\\
\ds \gamma\;=\;\ii q^{\hf}\;,\qquad
\kappa(x)\;=\;\frac{(qx,-q^2x;q^2)_\infty}{(q^2/x,-q^3/x;q^2)_\infty}\;,\\
\\
\ds \textrm{in general}\quad 
\frac{\kappa(x,q^{\hf}\gamma)}{\kappa(x,\gamma)}\;=\;
\frac{(q^2\gamma^2 x, q^2/\gamma^2x;q^2)_\infty}{(qx/\gamma^2,q^3\gamma^2/x;q^2)_\infty}\;.
\end{array}
\end{equation}
The model is physical for the real values of $q$, and its partition function $\mathbf{z}(x)$ per one weight $\Vop_x$ is the solution of the system of functional equations
\begin{equation}
\mathbf{z}(x)\mathbf{z}(q^2/x)\;=\;1\;,\quad
\frac{\mathbf{z}(x)}{\mathbf{z}(q/x)}\;=\;\kappa(x)\;,
\end{equation}
and it is given by 
\begin{equation}
\begin{array}{l}
\ds \gamma\;=\;\ii, \quad
\log\mathbf{z}(x)\;=\;-\sum_{n=1}^\infty \frac{q^{2n}(x^{2n}-q^{4n}/x^{2n})}{n(1-q^{4n})(1+q^{2n})}\;,\\
\\
\ds \gamma\;=\;\ii q^{\pm \hf}\;,\quad
\log\mathbf{z}(x)\;=\;-\sum_{n=1}^\infty \frac{q^n+(-q^2)^n}{n(1-q^{2n})(1+q^{n})} (x^n-q^{2n}/x^n)\;,\\
\\
\ds \textrm{in general}\quad
\frac{\mathbf{z}(x,q\gamma)}{\mathbf{z}(x,\gamma)}\;=\;
\frac{(q^2/\gamma^2 x, q^2\gamma^2 x;q^2)_\infty}{(x/\gamma^2,q^4\gamma^2/x;q^2)_\infty}\;.
\end{array}
\end{equation}
The partition function $\mathbf{z}_s$ per the central spin $\boldsymbol{S}_m$ 
is given by
\begin{equation}
\mathbf{z}_s\;=\;\frac{1}{\kappa(1)}\;=\;\kappa(q)\;.
\end{equation}
All partition functions belong to the class of the 8-vertex partition functions \cite{Baxbook}, and therefore the model is off-critical despite its ``rational'' nature similarly to the model described in \cite{descendant}.


\section{Modular representation.}\label{M-Section}

In this section we proceed to the modular representation of the Weyl algebras arisen in the homomorphisms (\ref{Hom1}) and (\ref{Hom2}). As the result, the hyperbolic Kashiwara-Miwa model will be systematically derived.

\subsection{Definition of the modular representations}

Faddeev's modular double\footnote{
Besides the Faddeev modular double, one can mention also the Dimofte-Gaiotto-Gukov double
\cite{DGG,Kash2} defined by
$$
\begin{array}{l}
\ds \uop^2|\theta,m\rangle = |\theta,m\rangle \EXP^{\ii\theta} q^{m},\quad
\vop | \theta,m\rangle = |\theta+\ii\hslash,m+1\rangle,\quad
\EXP^{-\hslash}=q,\quad \uop^2\vop=q^2\vop\uop^2,\\
\\
\ds \overline{\uop}^2|\theta,m\rangle = |\theta,m\rangle \EXP^{-\ii\theta} q^m,\quad
\overline{\vop}|\theta,m\rangle = |\theta-\ii\hslash,m+1\rangle,\quad \overline{\uop}^2\overline{\vop}=q^2\overline{\vop}\overline{\uop}^2\;,
\end{array}
$$
and the Andersen-Kashaev \cite{AndKash,K16} double defined by 
$$
\begin{array}{l}
\ds \uop^2|x,m\rangle = |x,m\rangle \omega \EXP^{2\pi\bk x},\;\;
\vop |x,m\rangle = |x+\ii\bk,m+1\rangle,\;\;
\uop^2\vop=\omega q^2 \vop\uop^2,\;\; q=\EXP^{\ii\pi\bk^2},\;\;\omega=\EXP^{2\pi\ii/N},\\
\\
\ds \overline{\uop}^2|x,m\rangle = |x,m\rangle \omega^{-1} \EXP^{2\pi\bk^{-1} x},\;\;
\overline{\vop} |x,m\rangle = |x-\ii\bk^{-1},m+1\rangle,\;\;
\overline{\uop}^2\overline{\vop}=\omega^{-1} \overline{q}^2 \overline{\vop}\overline{\uop}^2,\;\; q=\EXP^{-\ii\pi\bk^{-2}}.
\end{array}
$$
Each of these doubles can be used in our approach.
}
$\mathcal{W}_{q,\overline{q}}$ of the Weyl algebra is defined by 
\begin{equation}\label{genmod}
\uop |x\rangle = |x\rangle \EXP^{\pi\bk x},\;\;
\vop |x\rangle = |x+\ii\bk\rangle,\;\;
\overline{\uop} | x \rangle = | x \rangle \EXP^{\pi\bk^{-1}x},\quad
\overline{\vop} | x \rangle = | x - \ii\bk^{-1}\rangle,
\end{equation}
with 
\begin{equation}
q\;=\;\EXP^{\ii\pi\bk^2}\;,\quad \overline{q}\;=\;\EXP^{-\ii\pi\bk^{-2}}\;.
\end{equation}
Each shift $|x\rangle \to |x\pm\ii\bk\rangle$ is to be understood as a continuous shift of an integration contour of $\psi(x)=\langle \psi | x \rangle$ to the complex plane. This means in particular that the space of the wave functions $\psi(x)$ is not just $L_2$, it is also the space of functions analytical in a proper strip around the real axis.

The regime when the over-line stands for the complex conjugation, $\bk=\EXP^{\ii\theta}$, $\overline{\bk}=\bk^{-1}=\EXP^{-\ii\theta}$, is called the strongly coupling regime.

The modular double construction (\ref{genmod}) can be applied both to the homomorphism (\ref{Hom1}) and to the homomorphism (\ref{Hom2}). 
Through all this section we will use the following notations:
\begin{equation}\label{S}
S(x)\;=\;2\sinh \left(\pi\bk x\right) \sinh \left(\pi\bk^{-1}x\right)\;,
\end{equation}
and for the shortness 
\begin{equation}\label{M1}
\left\{
\begin{array}{l}
\ds u\;=\;\EXP^{\pi\bk x}\;,\quad
\overline{u}\;=\;\EXP^{\pi\bk^{\!-1} x}\;,\quad
q\;=\;\EXP^{\ii\pi\bk^2}\;,\quad 
\overline{q}\;=\;\EXP^{-\ii\pi\bk^{\!-2}}\;,\\
\\
\ds 
k = -q^{\hf} \EXP^{\pi\bk\sigma},\;\;
\overline{k} = -\overline{q}^{\hf}\EXP^{\pi\bk^{\!-1}\sigma},\;\;
z = \frac{q}{k^2} = \EXP^{-2\pi\bk \sigma},\;\;
\overline{z} = \frac{\overline{q}}{\overline{k}^2} = \EXP^{-2\pi\bk^{\!-1}\sigma}.
\end{array}\right.
\end{equation}
The modular representation associated with the homomorphism (\ref{Hom1}) is given by
\begin{equation}\label{MF}
\left\{
\begin{array}{l}
\ds \phi_{\omega,\lambda}(\mathcal{K})\, |\sigma\rangle \;=\; |\sigma\rangle\,k\,\EXP^{2\pi\bk\omega}\;,\quad 
\phi_{\omega,\lambda}(\mathcal{K}')\, |\sigma\rangle \;=\; |\sigma\rangle\,k\,\EXP^{-2\pi\bk\omega}\;,\\
\\
\ds \phi_{\omega,\lambda}(\mathcal{E}^{+})\, |\sigma\rangle \;=\; |\sigma+\ii\bk \rangle\,(-qk)\,\EXP^{-2\pi\bk\lambda}\;,\\
\\
\ds 
\phi_{\omega,\lambda}(\mathcal{E}^{-})\, |\sigma\rangle \;=\; |\sigma-\ii\bk \rangle\,
(q^{-1}k-k^{-1}) \, \EXP^{2\pi\bk\lambda}\;,
\end{array}\right.
\end{equation}
where all notations are fixed by (\ref{M1}), and the states are normalised by 
\begin{equation}\label{MFn}
\langle \sigma | \sigma' \rangle \;=\; \delta(\sigma-\sigma') (z;q^2)_\infty (\overline{z};\overline{q}^2)_\infty\;.
\end{equation}
Later we will see that this is the most convenient normalisation here.
The modular partner to the representation (\ref{MF}) can be obtained from (\ref{MF}) by a simple complex conjugation. In its turn, the modular representation for the homomorphism (\ref{Hom2}) is defined by 
\begin{equation}\label{MV}
\left\{
\begin{array}{l}
\ds \pi_{\omega,\mu}(\mathcal{K})\, | x \rangle \;=\; \biggl( |x+\ii\bk\rangle - |x-\ii\bk\rangle\biggr)\,\EXP^{2\pi\bk\omega}/S(x)\;,\\
\\
\pi_{\omega,\mu}(\mathcal{E}^{+})\, | x \rangle \;=\; 
\biggl( |x-\ii\bk\rangle\, u  - |x+\ii\bk\rangle\, u^{-1} \biggr)\, \EXP^{-2\pi\bk\mu}/S(x)\;,\\
\\
\pi_{\omega,\mu}(\mathcal{E}^{-})\, | x \rangle \;=\; 
\biggl( |x+\ii\bk\rangle\, u  - |x-\ii\bk\rangle\, u^{-1} \biggr)\,\EXP^{2\pi\bk\mu} /S(x)\;,
\end{array}\right.
\end{equation}
with the normalisation 
\begin{equation}\label{MVn}
\langle x | x' \rangle \;=\; \frac{1}{2S(x)}\; \biggl( \delta(x-x') + \delta(x+x')\biggr)\;.
\end{equation}

\subsection{Equivalence of the modular representations.}

Two modular representations, (\ref{MF}) and (\ref{MV}), are equivalent. This statement follows from the completeness and non-generativity of the matrix elements $\langle \lambda,\sigma|\mu, x\rangle$.

Before we write down these matrix elements, we need some notations and definitions.
Let 
\begin{equation}\label{MG}
G(x)\;=\;
\frac{\EXP^{-\frac{\ii\pi}{8}+\frac{\ii\pi x^2}{2}}}{ \bk^{\hf} q^{^1\!\!/\!_4} \BH(u)}\;,
\end{equation}
and let further 
\begin{equation}
\Chi(u)=\Chi(u,z;q)\;,\quad \overline{\Chi(u)}\;=\;\Chi(\overline{u},\overline{z};\overline{q})
\end{equation}
are given by (\ref{Bchi}) in Appendix \ref{Chi-Section}. Let also 
\begin{equation}\label{MPsi}
\Psi(\sigma,x)\;=\;G(x)\,
\biggl(\EXP^{\ii\pi\sigma x} \overline{\Chi(1/u)}\Chi(u) - \EXP^{-\ii\pi\sigma x} \overline{\Chi(u)}\Chi(1/u)\biggr)\;.
\end{equation}
\begin{proposition}\label{prop9.1}
The problem of the construction of matrix elements between the representations (\ref{MF},\ref{MFn}) and (\ref{MV},\ref{MVn}) has unique solution,
\begin{equation}\label{Bsol}
\langle \sigma | x \rangle \;=\; \EXP^{2\pi\ii(\lambda-\mu)\sigma}\;\Psi(\sigma,x)\;,\;\;
\langle x | \sigma \rangle \;=\; N^{-1}\, \EXP^{2\pi\ii(\mu-\lambda)\sigma}\;\Psi(\sigma,x)\;,
\end{equation}
where
\begin{equation}
N\;=\;\frac{2}{(q\overline{q})^{^1\!/\!_4} (q^2;q^2)_\infty (\overline{q}^2;\overline{q}^2)_\infty}\;.
\end{equation}
\end{proposition} 
\noindent\textbf{Sketch proof.} 
How one can derive (\ref{Bsol}). First, one has to consider the equations for $\mathcal{K}$,
\begin{equation}\label{Bev}
\left\{
\begin{array}{l}
\ds \langle \sigma | \mathcal{K}_0 | x \rangle \;=\; 
\frac{\langle \sigma | x+\ii\bk\rangle - \langle \sigma | x-\ii\bk\rangle}{2\sinh (\pi \bk x)}\;=\; -q^{\hf}\EXP^{\pi\bk \sigma} 
\langle \sigma | x \rangle\;,\\
\\
\ds \langle \sigma | \overline{\mathcal{K}}_0 | x \rangle \;=\; 
\frac{\langle \sigma | x-\ii\bk^{-1}\rangle - \langle \sigma | x+\ii\bk^{-1}\rangle}{2\sinh (\pi \bk^{-1} x)}\;=\; -\overline{q}^{\hf}\EXP^{\pi\bk^{-1} \sigma} 
\langle \sigma | x \rangle\;.
\end{array}\right.
\end{equation}
Essential condition here is that $\langle \sigma | x \rangle$ must be an entire function in the whole complex plane of $x$ (and, as the consequence -- in whole plane $\sigma$). The method of construction of such solutions has been described in e.g.\cite{Sergeev_2005,Kashaev_2018}. The method is based on a selection of holomorphic and anti-holomorphic parts, on their Wronskian and on cancellation of poles. In our case $\Chi(u)$ is the holomorphic solution of (\ref{Bchieq}), the theta-function $\BH(u)$ in the denominator of (\ref{MG}) is the Wronskian (\ref{BWr}), and the cancellation of the poles in the whole expression (\ref{Bsol}) follows from the Wronskian (\ref{BWr}),
\begin{equation}
\frac{\Chi(q^{-m})}{\Chi(q^m)}\;=\;z^m\;.
\end{equation}
Also one can note, the function $\Psi(\sigma, x)$ is real in the strong coupling regime, what follows from (\ref{Ajac}). Next, when the matrix elements $\langle x | \sigma\rangle$ are obtained, one can check straightforwardly that $\mathcal{E}^{\pm}$, acting on the left in $\langle \sigma |\mathcal{E}^{\pm} | x \rangle$, reproduce the representation (\ref{MF}). \hfill $\square$
\\

\noindent\textbf{Comment}. 
The normalisation (\ref{MFn}) contains the second order zeros on the real axis,\begin{equation}
\textrm{Second order zeros:}\quad \sigma=\ii ( \bk - \bk^{-1}) n\;,\quad n\geq 0\;.
\end{equation}
At these points $\Psi(\sigma,x) = 0 $ as well. One can verify, the l'Hospitale rule works perfectly at these points, so there are no singular points in the spectrum of $\mathcal{K}$ and therefore representations (\ref{MF}) and (\ref{MV}) are equivalent.

\subsection{Co-product.}

The co-product structure is completely described in Section \ref{V-Section}, the modular case corresponds to $A=1$, $B=-1$ and $\varepsilon=-1$.

\subsection{Intertwining operators $V$, $\overline{V}$ and the Star-Triangle relation.}

Section \ref{V-Section} contains all required relations, e.g. the relations (\ref{req}). 
They just could be solved in the terms of Faddeev's dilogarithm.

Faddeev's dilogarithm is the uniquely defined solution of the difference equations
\begin{equation}
\frac{\varphi(x-\ii\bk^{\pm 1}/2)}{\varphi(x+\ii\bk^{\pm 1}/2)}\;=\;
1 + \EXP^{2\pi x \bk^{\pm 1}}\;,
\end{equation}
analytical in the strip $-\textrm{Re}(\bk) < \textrm{Im}(x) < \textrm{Re}(\bk)$. This solution can be presented as 
\begin{equation}
\varphi(x)\;=\;\exp\biggl( \frac{1}{4}\int_{\mathbb{R}+\ii 0} \frac{\EXP^{-2\ii xw}}{\sinh(w\bk) \sinh(w/\bk)}\, \frac{dw}{w} \biggr)\;=\;
\frac{(\EXP^{2\pi (z+\eta)\bk};q^2)_\infty}{(\EXP^{2\pi(z-\eta)/\bk};\overline{q}^2)_\infty}\;,
\end{equation}
where
\begin{equation}
\eta\;=\;\ii\;(\bk+\bk^{-1})/2\;.
\end{equation}
Thus, the difference equations (\ref{req}) and their modular partners for the function
\begin{equation}\label{V1}
\langle \mu, x | \mu', y \rangle \;=\; V_{\mu-\mu'}(x,y)
\end{equation}
have the following solution:
\begin{equation}\label{V2}
V_{\mu}(x,y)\;=\;\frac{1}{\Phi(\mu)}\,\EXP^{2\pi\ii(\mu-\eta)x}
\,
\frac{\varphi(\frac{x-y}{2}-\mu+\eta)}{\varphi(\frac{x-y}{2}+\mu-\eta)}\,
\frac{\varphi(\frac{x+y}{2}-\mu+\eta)}{\varphi(\frac{x+y}{2}+\mu-\eta)}\;.
\end{equation}
The same expression, together with the scalar multiplier $\Phi(\mu)$, comes from the integral 
\begin{equation}
V_\mu(x,y)\;=\;N^{-1}\int d\sigma \, \EXP^{2\pi\ii\mu\sigma}\, 
\frac{\Psi(\sigma,x)\Psi(\sigma,y)}{(z;q^2)_\infty (\overline{z};\overline{q}^2)_\infty}\;.
\end{equation}
The value of $\Phi(\mu)$ has been obtained numerically,
\begin{equation}\label{B50}
\Phi(\mu)\;=\;2\gamma\phi_0^2 \EXP^{-2\ii\pi\mu^2-2\pi\bk\mu}\;
\varphi(\eta-2\mu)\;,
\end{equation}
where
\begin{equation}
\phi_0^2\;=\;(q/\overline{q})^{^1\!/\!_{12}}\;=\;\EXP^{-\ii\pi\eta^2/6-\ii\pi/12}\;,\;\;
\gamma\;=\;\EXP^{\ii\pi/4}\;.
\end{equation}
It must be noted, the modular symmetry seems to be broken in the exponent in (\ref{B50}). The reason for this is that the symmetry already has been broken on the level of the definition of $k\;=\;-\ii \EXP^{\pi\bk(\sigma+\ii(\bk-\bk^{\!-1})/2)}$ and
$\overline{k}\;=\;\ii \EXP^{\pi\bk^{\!-1}(\sigma+\ii(\bk-\bk^{\!-1})/2)}$.
The function (\ref{V2}) is normalised according to (\ref{MVn}),
\begin{equation}
V_{\ii 0}(x,y)\;=\;\frac{1}{2 S(x)} \, \biggl(\delta(x-y)+\delta(x+y)\biggr)\;.
\end{equation}
Also, the function $V$ is symmetric,
\begin{equation}
V_\mu(x,y)\;=\;V_\mu(\pm x,\pm y)\;=\;V_\mu(\pm y,\pm x)\;,
\end{equation}
and it satisfies the summation formula,
\begin{equation}\label{M-sum}
\int dy V_{\mu}(x,y)\, S(y)\, V_{\mu'}(y,z)\;=\; V_{\mu+\mu'}(x,z)\;.
\end{equation}
The weight $\overline{V}$ for the Star-Triangle relation is now
\begin{equation}\label{V3}
\overline{V}_{\mu}(x,y)\;=\;V_{\eta-\mu}(x,y)\;.
\end{equation}
We do not discuss in details the Star-Triangle relation and corresponding integrable system here since all relevant details (except the summation formula) can be found in \cite{BSKels}.

\section{Summary and conclusion}

This paper is an experience of the Hopf algebra approach to the generalised $q$-oscillator algebra. We have shown that the $R$-matrices for the rational (and hyperbolic) Kashiwara-Miwa models appear as the canonical intertwining operator for the co-product of $q$-oscillators. This is our main result.

Another remarkable observation made in this paper is the collection of the summation formulas (\ref{F-sum},\ref{V-sum},\ref{M-sum}) for the Boltzmann weights of the integrable models.

Only two representations are discussed in the paper. They are the Fock space representation and Faddeev's modular double representation. Other representations and other modular doubles we leave for further publications.

Our final remark is that the homomorphism (\ref{Hom2}) appeared in the theory of the Tetrahedron equation \cite{BMS08}. Potentially, the Hopf algebra structure of the $q$-oscillators may have an impact to the three-dimensional integrable theories.
\\

\noindent
\textbf{Acknowledgements.} The author would like to thank R. Kashaev, V. Bazhanov and V. Mangazeev for valuable discussions. 


\appendix

\bigskip

\begin{center}
    {\bf APPENDIX}
\end{center}

\section{General notations used in this paper.}\label{A1-Section}

In this paper various Pochhammer symbols have been used:
\begin{equation}
(u;q^2)_n\;=\;\prod_{k=0}^{n-1} (1-xq^{2n})\;,\quad
(q^2;q^2)_{a,b,\dots}\;=\;(q^2;q^2)_a (q^2;q^2)_b\cdots
\end{equation}
\begin{equation}
(x,y,\cdots;q^2)_\infty\;=\;(x;q^2)_\infty (y;q^2)_\infty \cdots
\end{equation}
A symbol in the square brackets implies
\begin{equation}
[z]\;=\;z-z^{-1}\;.
\end{equation}
Also various theta-functions have been used here:
\begin{equation}\label{ATTH}
\begin{array}{l}
\ds \theta(u)=\theta_4(u)=\sum_{n\in\mathbb{Z}} q^{n^2/2} (-u)^n=(q^{\hf} u,q^{\hf}u^{-1},q;q)_\infty\;,\quad
\theta_3(u)=\theta_4(-u)\;,\\
\\
\ds \Theta(u)=\Theta_4(u)=\sum_{n\in\mathbb{Z}} q^{n^2} (-u)^n = (qu,qu^{-1},q^2;q^2)_\infty\;,\quad
\Theta_3(u)=\Theta_4(-u)\;,\\
\\
\ds \BH(u)\;=\;\sum_{n\in\mathbb{Z}} q^{n(n-1)} (-)^n u^{2n-1}\;=\;u^{-1}(u^2,q^2u^{-2},q^2;q^2)_\infty
\end{array}
\end{equation}
The Jacobi transform for $\BH(u)$ was also used,
\begin{equation}\label{Ajac}
q^{^1\!\!/\!_4} \frac{(u^2,q^2/u^2,q^2;q^2)_\infty}{u}\;=\;
\EXP^{\frac{3\pi\ii}{4}+\ii\pi x^2} \bk^{\!-1}\,
\overline{q}^{^1\!\!/\!_4} \frac{(\overline{u}^2,\overline{q}^2/\overline{u}^2,\overline{q}^2;\overline{q}^2)_\infty}{\overline{u}}\;.
\end{equation}
The theta-constants appear in the normalisation constants, e.g. in (\ref{NT4}),
\begin{equation}\label{AT4}
\Theta_4\;=\;\Theta_4(1)\;=\;(q,q,q^2;q^2)\;=\;\sum_{n\in\mathbb{Z}} (-)^n q^{n^2}\;=\;
\frac{(q;q)_\infty}{(-q;q)_\infty}\;,
\end{equation}
The coefficient in (\ref{F-M}) is
\begin{equation}\label{AFM}
M_F\;=\;-q^{-\hf} \sum_{m\in\mathbb{Z}} (-)^m q^{m(3m+1)/2}\;=\;-q^{-\hf} (q;q)_\infty\;.
\end{equation}
The normalisation (\ref{FKGn}) is based on 
\begin{equation}\label{AFnorm}
\begin{array}{l}
\ds \sum_{n_1,n_2=0}^\infty \frac{q^{2n_1n_2}}{(q^2;q^2)_{n_1,n_2}}z_1^{n_1}z_2^{n_2}
\;=\;\frac{(z_1z_2;q^2)_\infty}{(z_1,z_2;q^2)_\infty}\;,\\
\\
\ds 
\sum_{n_1,n_2=0}^\infty \frac{q^{2n_1n_2}}{(q^2;q^2)_{n_1,n_2}}\EXP^{\ii\varphi (n_1-n_2)} \;=\; 
\frac{2\pi\delta(\varphi)}{(q^2;q^2)_\infty}\;,\\
\\
\ds 
\sum_{n=0}^\infty \frac{q^{2n^2}}{(q^2;q^2)_n^2} \;=\; 
\frac{1}{(q^2;q^2)_\infty},
\end{array}
\end{equation}
Derivation of these identities is based on the Gauss identity,
\begin{equation}
\sum_{n=0}^\infty x^n \frac{(z;q)_n}{(q;q)_n}\;=\;
\frac{(xz;q)_\infty}{(x;q)_\infty}\;.
\end{equation}

\section{Polynomials $\pol_a(\xi)$ and their properties.}\label{A2-Section}

Polynomials $\pol_a(\xi)$ are defined by 
\begin{equation}\label{A1}
\pol_n(\xi)\;=\;\sum_{k=0}^n \xi^{2k-n} q^{-2k(n-k)} \left(\!\!\begin{array}{c} n \\ k \end{array}\!\!\right)_{\!q^2}\;.
\end{equation}
These polynomials appear as the solution of the first recursion,
\begin{equation}\label{A2}
\pol_{n+1}(\xi) \;=\; q^{-n} \, 
\frac{\xi^2 \pol_n(q\xi) - \xi^{-2} \pol_n(\xi/q)}{\xi-\xi^{-1}}\;,\quad \pol_0(\xi)\;=\;1\;,
\end{equation}
or, as the solution of the second recursion,
\begin{equation}\label{A4}
\pol_{n+1}(\xi)  \;=\; (\xi+\xi^{-1}) \pol_n(\xi) - (1-q^{-2n}) \pol_{n-1}(\xi)\;.
\end{equation}
Polynomials $\pol_a(\xi)$ satisfy the homogeneous difference equations:
\begin{equation}\label{A3}
\pol_n(\xi)\;=\;q^{-n}\,
\frac{\xi \pol_n(q\xi) - \xi^{-1} \pol_n(\xi/q)}{\xi-\xi^{-1}}\;.
\end{equation}
The consequence of the recursion and difference relations is the backward recursion:
\begin{equation}\label{A5}
(1-q^{-2n}) \pol_{n-1}(\xi)\;=\;q^{-n} \, \frac{\pol_n(q\xi) - \pol_n(\xi/q)}{\xi-\xi^{-1}}\;.
\end{equation}
Relations (\ref{A2},\ref{A3},\ref{A5}) correspond to the action of the algebra in the representation (\ref{PsiF}). The polynomials $\pol_a(\xi)$ have two generating functions:
\begin{equation}\label{A6}
\sum_{n=0}^\infty (-)^n q^{n(n-1)/2} \frac{z^n}{(q;q)_n} \; \pol_n(\xi)\;=\;
\frac{(z\xi,z/\xi;q)_\infty}{(z^2/q;q^2)_\infty}\;.
\end{equation}
and
\begin{equation}\label{A7}
\sum_{n=0}^\infty (-)^n q^{n(n-1)} \frac{z^n}{(q^2;q^2)_n} \; \pol_n(u) \pol_n(v)\;=\;
\frac{(zuv,z/uv,zu/v,zv/u;q^2)_\infty}{(z^2/q^2;q^2)_\infty}\;.
\end{equation}
In particular, 
\begin{equation}\label{A8}
\sum_{n=0}^\infty (-)^n q^{n(n+1)} \frac{\pol_n(q^{m+\hf}) \pol_n(q^{m'+\hf})}{(q^2;q^2)_n} 
\;=\;
\delta_{m,m'} (-)^{m} \frac{q^{-m^2}}{1-q^{2m+1}} \Theta_4\;,
\end{equation}
where $m,m'\geq 0$. These formulas are related to the normalisation (\ref{Fnorm}) and to the weights (\ref{F-V-1}) and (\ref{V-V-1}).

\section{Function $\Chi$ and its properties}\label{Chi-Section}

Define $\Chi(u,z)$ by
\begin{equation}\label{Bchi}
\Chi(u,z;q)\;=\;\sum_{n=0}^\infty (-)^n q^{n(n+1)} \frac{(z;q^2)_n}{(q^2;q^2)_n} u^{2n}\;.
\end{equation}
In particular, if 
\begin{equation}
z\;=\;q^{-2m}\;,\quad\textrm{then}\quad \Chi(u,q^{-2m})\;=\;u^m \pol_m(u)\;,
\end{equation}
so that $\Chi(u,z)$ is a generalisation of the polynomials $\pol_m(u)$. Function $\Chi$ satisfies an equation analogues to (\ref{A3}), and this equation is related to с (\ref{Bev}):
\begin{equation}\label{Bchieq}
\Chi(u/q,z)\;=\;(1-u^2)\Chi(u,z) + zu^2 \Chi(qu,z)\;.
\end{equation}
Relations (\ref{A2}) and (\ref{A5}) have the analogues
\begin{equation}
\begin{array}{l}
\ds (1-z) (1-u^2) \Chi(u,q^2z) \;=\; \Chi(q^{-1}u,z) - z \Chi(qu,z)\;,\\
\\
\ds (1-u^2) \Chi(u,q^{-2}z)\;=\;\Chi(q^{-1}u,z) - z u^4 \Chi(qu,z)\;.
\end{array}
\end{equation}
Analogue of (\ref{A4}) is
\begin{equation}
u^{-1}\Chi(u,q^{-2}z) + u (1-z) \Chi(u,q^2z) \;=\; (u+u^{-1}) \Chi(u,z)\;.
\end{equation}
All these equations are related to Proposition \ref{prop9.1}. The Wronskian of $\Chi$ is 
\begin{equation}\label{BWr}
\Chi(u/q,z) \Chi(1/u,z) - z \Chi(u,z)\Chi(q/u,z)\;=\;(u^2,q^2/u^2,z;q^2)_\infty\;.
\end{equation}

\section{Additional representations of the $q$-oscillator.}\label{Add-Section}

We have discussed only the Fock space representation and one of the possible modular representations of the $q$-oscillator in the main text. It worth to mention some other representations and their ``multiplication table'' in the appendix.

\subsection{Representation $\widetilde{\mathbb{F}}$.}

The representation $\widetilde{\mathbb{F}}$, called sometimes the ``anti-Fock'' representation, can be described as a continuation of the usual Fock space (\ref{MyFock}) to the negative occupation numbers. As well, it can be described as the result of the change $q\to q^{-1}$ in the representation (\ref{MyFock}):
\begin{equation}\label{MyF2}
\begin{array}{l}
\ds
\widetilde{\phi}_{\omega,\lambda}(\mathcal{E}^{-})\, | a \rangle = | a+1 \rangle\,\lambda\, \frac{\sqrt{1-q^{2a+2}}}{q^{a+1}}, \;\;
\widetilde{\phi}_{\omega,\lambda}(\mathcal{K}) \,| a \rangle = | a \rangle \,\omega\, q^{-a-\hf},\\
\\
\ds 
\widetilde{\phi}_{\omega,\lambda}(\mathcal{K}') \,| a \rangle = | a \rangle \,\omega^{-1}\, q^{-a-\hf},\;\;
\widetilde{\phi}_{\omega,\lambda}(\mathcal{E}^{+})\,  | a \rangle = | a-1 \rangle \, (-\lambda)^{-1} \frac{\sqrt{1-q^{2a}}}{q^a},
\end{array}
\end{equation}
Constructing the $\mathbb{V}$-form of this representation, one will observe that $\mathbb{S}\;=\;\mathbb{T}$ -- the circle. In its turn, constructing the co-product, one obtains $\omega=\pm q^{2k}$, $k\in\mathbb{Z}$, what means
\begin{equation}
\widetilde{\mathbb{F}}\otimes\widetilde{\mathbb{F}}\;=\;\left( \mathbb{Z}_2\otimes\mathbb{Z}\right)
\otimes \widetilde{\mathbb{F}}\;.
\end{equation}

Next, when one considers the product $\mathbb{F}\otimes\widetilde{\mathbb{F}}$, when the method of $\mathbb{V}$-form can not be applied, something different happens:
\begin{equation}\label{Add-FnF}
\mathbb{F}\otimes\widetilde{\mathbb{F}}\;=\;\sum_{n=0}^\infty \mathbb{N}_n\;.
\end{equation}
Here the space $\mathbb{N}_0$ is defined as follows:
\begin{equation}
|\!|\psi_0\rangle=|0\rangle\otimes |0\rangle\;,\quad
|\!|\psi_n\rangle= \Delta(\mathcal{K}')\,|\!|\psi_0\rangle\;,\quad
\Delta(\mathcal{K})\,|\!|\psi_n\rangle  =0\;.
\end{equation}
In this basis 
\begin{equation}
\Delta(\mathcal{E}^{\pm})\, |\!|\psi_n\rangle = |\!|\psi_n\rangle \left(-\frac{\omega_1}{\omega_2}\right)^{\mp 1} \, q^{\pm n}\;.
\end{equation}
The other irreducible representations in (\ref{Add-FnF}) are defined as the factor spaces 
\begin{equation}
\mathbb{N}_n\;=\;\biggl\{ |\!| v \rangle \;:\;\; \Delta(\mathcal{K})^{n+1}\, |\!| v \rangle = 0 \biggr\} /
\biggl\{ |\!| v \rangle \;:\;\; \Delta(\mathcal{K})^n \, |\!| v \rangle = 0\biggr\}\;.
\end{equation}
Also, one can select the set of irreducible $(n+1)\times (n+1)$-dimensional representations of $\mathcal{O}_q$ inside the product (\ref{Add-FnF}):
\begin{equation}
\mathbb{J}_n\;=\;\textrm{Span}\biggl( |a\rangle\otimes |b\rangle\;,\;\; a+b=n\;,\;\; n\geq 0\biggr)\;,
\quad \Delta(\mathcal{E}^{\pm})\;:\;\;\mathbb{J}_n\;\to\;\mathbb{J}_n\;.
\end{equation}
In these finite-dimensional representations the matrices $\Delta(\mathcal{KK}')$ are nilpotent Jordan cells.

\subsection{Representation $\mathbb{A}$.}

There is one more representation of the $q$-oscillator to be mentioned.
This representation corresponds to the homomorphism (\ref{Hom2}) with unitary $\uop$. It can be defined e.g. as follows:
\begin{equation}\label{W2}
\begin{array}{l}
\ds \mathcal{K}_0\, | \alpha \rangle = |\alpha-1\rangle,\\
\\
\ds 
\mathcal{E}^{+} | \alpha \rangle = \biggl(| \alpha \rangle + |\alpha-1\rangle q^{-\hf} \biggr) q^\alpha,\\
\\
\ds 
\mathcal{E}^{-} | \alpha \rangle = \biggl( | \alpha \rangle - |\alpha-1\rangle q^{\hf} \biggr) q^{-\alpha}.
\end{array}
\end{equation}
Here $\alpha\in\mathbb{Z}$. For the $\mathbb{V}$-form of this representation, see (\ref{V}),
\begin{equation}
\mathbb{V}\;=\;\textrm{Span}\biggr(|\mu,\xi\rangle\;,\;\;\xi\;=\;-\mu q^{m-\hf}\;,\;\;m\in\mathbb{Z}\biggr)\;,
\end{equation}
while the additional conditions for $|\psi\rangle\in\mathbb{A}$ are of the form
\begin{equation}\label{Acond}
\lim_{m\to - \infty} \frac{\langle \mu, m | \psi \rangle}{\mu^m q^{m(m-1)/2}}\;\to\; 0\;,\quad
\lim_{m\to + \infty} \frac{\langle \psi | \mu, m\rangle}{\mu^m q^{m(m-1)/2}}\;\to\; 0\;.
\end{equation}
For the co-product one obtains $\omega=q^k$, what gives
\begin{equation}
\mathbb{A}\otimes\mathbb{A}\;=\;\mathbb{Z}\otimes\mathbb{A}\;.
\end{equation}
In addition, it is not difficult to verify
\begin{equation}
\mathbb{F}\otimes\mathbb{A}\;=\;\mathbb{Z}\otimes\mathbb{F}\;.
\end{equation}
Since in all these examples $\omega$ is discrete, then the discrete spectral parameters appear in the corresponding Star-Triangle equations. For this reason all these examples are not included into the main text. 

\end{document}